\documentclass[]{aa}
\usepackage{graphicx}
\usepackage{longtable}

\def\ndeprime{$96$}
\def\nfit{$71$}
\def\nlow{$1109$}

\newcommand{\ind}[1]{_{\mathrm{#1}}}
\newcommand{\diff}{\mathrm{d}}
\newcommand{\mV}{m\ind{V}}

\newcommand\Dnu{\Delta\nu}
\newcommand\Tg{\Delta\Pi_1}

\newcommand\Teff{T\ind{eff}}

\newcommand\dnurot{\delta\nu\ind{rot}}
\newcommand\Vdip{V_1^2}
\newcommand\Vgeo{\tilde V_1^2}
\newcommand\vm{v_{\nm}}

\newcommand{\eva}{{\mathcal{E}}}
\newcommand{\cavp}{{\mathcal{A}}}
\newcommand{\dens}{{\mathcal{N}}}

\newcommand\numax{\nu\ind{max}}

\newcommand\np{{n\ind{p}}}

\newcommand\nm{{n\ind{m}}}

\newcommand{\BV}{Brunt-V\"ais\"al\"a}

\newcommand\Kepler{\emph{Kepler}}

\begin{document}

\title{Dipole modes with depressed amplitudes in red giants are mixed modes}
\titlerunning{Mixed modes and low visibility}
\authorrunning{Mosser et al.}
\author{%
 B. Mosser\inst{1},
 K. Belkacem\inst{1},
 C. Pin\c con\inst{1},
 M. Takata\inst{2},\\
 M. Vrard\inst{3},
 C. Barban\inst{1},
 M-J. Goupil\inst{1},
 T. Kallinger\inst{4},
 R. Samadi\inst{1}\\
 }

\institute{
 LESIA, Observatoire de Paris, PSL Research University, CNRS, Universit\'e Pierre et Marie Curie,
 Universit\'e Paris Diderot,  92195 Meudon, France
 \texttt{benoit.mosser@obspm.fr}
 \and
 Department of Astronomy, School of Science, The University of Tokyo, 7-3-1 Hongo, Bunkyo-ku, Tokyo 113-0033, Japan
 \and
 Instituto de Astrof\'isica e Ci\^encias do Espa\c co, Universidade do Porto, CAUP, Rua das Estrelas, 4150-762 Porto, Portugal
 \and
 Institute of Astrophysics, University of Vienna, T\"urkenschanzstrasse 17, Vienna 1180, Austria
}


\abstract{Seismic observations with the space-borne \Kepler\
mission have shown that a number of evolved stars exhibit
low-amplitude dipole modes, which are referred to as depressed
modes. Recently, these low amplitudes have been attributed to the
presence of a strong magnetic field in the stellar core of those
stars.
}
{We intend to study the properties of depressed modes in evolved
stars, which is a necessary condition before concluding on the
physical nature of the mechanism responsible for the reduction of
the dipole mode amplitudes.}
{We perform a thorough characterization of the global seismic
parameters of depressed dipole modes and show that these modes
have a mixed character. The observation of stars showing dipole
mixed modes that are depressed is especially useful for deriving
model-independent conclusions on the dipole mode damping.
}
{Observations prove that depressed dipole modes in red giants are
not pure pressure modes but mixed modes. This result invalidates
the hypothesis that the depressed dipole modes result from the
suppression of the oscillation in the radiative core of the stars.
Observations also show that, except for the visibility, the
seismic properties of the stars with depressed modes are
equivalent to those of normal stars.
}
{The mixed nature of the depressed modes in red giants and their
unperturbed global seismic parameters carry strong constraints on
the physical mechanism responsible for the damping of the
oscillation in the core. This mechanism is able to damp the
oscillation in the core but cannot fully suppress it. Moreover, it
cannot modify the radiative cavity probed by the gravity component
of the mixed modes. The recent mechanism involving high magnetic
field proposed for explaining depressed modes is not compliant
with the observations and cannot be used to infer the strength and
the prevalence of high magnetic fields in red giants.}


\maketitle

\section{Introduction}

Asteroseismic observations by the CoRoT and \Kepler\ space-borne
missions have provided new insights in stellar and Galactic
physics
\citep[e.g.,][]{2008Sci...322..558M,2009A&A...503L..21M,2011Sci...332..213C}.
The ability to derive fundamental stellar parameters such as
masses and radii over a wide range of stellar evolutionary states
from the main sequence to the asymptotic giant branch is certainly
one of the strongest impacts \citep[e.g.,][]{2010A&A...522A...1K}.
The rich nature of the red giant oscillation spectrum was largely
unexpected \citep{2009Natur.459..398D,2016cole.book..197M}. Red
giant asteroseismology has been boosted by the observation of
dipole modes that, due to their mixed nature, probe the stellar
core. They behave as gravity modes in the core and as pressure
modes in the envelope. The pressure components carry information
on the mass and radius of stars
\citep[e.g.,][]{2013A&A...550A.126M}, while the gravity components
of these mixed modes are directly sensitive to the size and mass
of the helium core
\citep{2013EPJWC..4303002M,2016MNRAS.457L..59L}, to the
evolutionary stage differing by the nuclear reaction at work
\citep{2011Natur.471..608B,2011A&A...532A..86M}, and to the mean
core rotation
\citep{2012Natur.481...55B,2012A&A...548A..10M,2012ApJ...756...19D,2014A&A...564A..27D}.

Due to the homology of the red giant interior structure, red giant
seismology is characterized by the many scaling relations between
global seismic parameters along stellar evolution
\citep[e.g.,][]{2009MNRAS.400L..80S,2010A&A...517A..22M,2011ApJ...741..119M,2012A&A...541A..51K}.
Exceptions to these relations are rare and most often explained by
specific features, as for instance the damping of the oscillation
in close binaries \citep{2014ApJ...785....5G} or very low
metallicity \citep{2014ApJ...785L..28E}. However, observations
have revealed that a family of red giants exhibit peculiar dipole
modes with low amplitudes \citep{2012A&A...537A..30M}. In some
extreme cases, dipole modes are not even detectable. Consequently,
these dipole modes have been called \emph{depressed
modes}\footnote{In the following, we use the term \emph{depressed}
for modes exhibiting diminished visibilities and keep the term
\emph{suppressed} for the suppression of the oscillation in the
stellar core. As shown in Section \ref{full_suppression}, the full
suppression of the oscillation in the core induces the mode
depression. In contrast, \emph{normal} modes have normal
visibilities. Suppression implicitly means full suppression, so
that we introduce the term \emph{partial suppression} when we have
to stress that the supposed suppression is not total. We do not
use the term \emph{mode suppression}, since it can only correspond
to a null visibility.}.

The first in-depth study of a star with depressed modes could not
explain this phenomenon \citep{2014A&A...563A..84G}. Then,
\cite{2015Sci...350..423F} addressed this issue by using a twofold
approach. First, the authors expressed the dipole mode
visibilities in the limit of full suppression of the oscillation
in the red giant core. In other words, they assumed that the mode
energy that leaks in the radiative interior of red giants is
totally lost. As a consequence, dipole modes are no longer mixed
modes but only lie in the upper (acoustic) cavity of red giants.
This assumption is validated by the authors by means of a
comparison between observed and computed mode visibilities.
Second, it is conjectured that the extra loss of the mode energy
is caused by a strong magnetic field, which scatters waves leaking
in the core. This prevents them from constructing a standing wave
in the inner resonant cavity of those stars. This has been named
as the \emph{magnetic greenhouse effect}. The angular degree
dependence of the energy leakage has been verified with the
quadrupole modes \citep{2016PASA...33...11S}. This appealing
scenario has then been taken as granted by
\cite{2016Natur.529..364S} and \cite{2016ApJ...824...14C} to infer
the prevalence of magnetic fields in the core of oscillating red
giant stars observed by \emph{Kepler}.

However, before firmly concluding about the presence of a magnetic
field in the core of red giants with depressed modes, the
hypothesis of the suppression of the oscillation in the core has
to be validated. In fact, the nature of the physical mechanism
responsible for the reduction of dipole mode visibilities is not
directly inferred by the observation of low mode visibilities.
Therefore, the identification of a magnetic field as the physical
mechanism responsible for the suppression of the oscillation in
the core of those stars demands further direct observational
confirmations. A strong magnetic field is a possible solution, but
not the only one. \cite{2015Sci...350..423F} note that rapid core
rotation should have the same effect, but would require much
larger rotation rates than observed in red giants
\citep{2012A&A...548A..10M,2014A&A...564A..27D}. Indeed, any
strong damping in the core such as radiative damping or gravity
wave reflection, caused for instance by a steep composition
gradient above the hydrogen burning shell, could also explain the
depressed modes \citep{2012A&A...539A..83D}.

In this work, we aim at providing a complete characterization of
the population of red giants with depressed modes, using
\emph{Kepler} observations. This study is motivated by the
observation of stars showing dipole modes that are both mixed and
depressed. We argue that the full characterization of stars with
depressed visibilities can provide strong constraints on the
physical mechanism responsible for the damping of the oscillation.

The article is organized as follows. Section \ref{formalism}
presents the theoretical background of mode visibility, with an
emphasis on the distinction between full and partial mode damping
in the core of red giants. In Section~\ref{observations}, we
undertake the characterization of depressed modes with a mixed
character, hereafter named \emph{depressed mixed modes}. We first
explain how they were identified, then exploit their observations
with the determination of their global seismic properties. We also
consider stars where, owing to the too low visibility of dipole
modes, the gravity-dominated mixed modes are apparently absent.
Global properties of stars are then used to test the depressed
visibilities predicted when the oscillation is suppressed in the
stellar core (Section \ref{modele}). Section \ref{discussion} is
devoted to discussion, with particular attention to the nature of
the mechanism responsible for the extra-damping of depressed
modes. Finally, Section~\ref{conclusion} is dedicated to
conclusions.

\section{Dipole mode visibilities\label{formalism}}

\subsection{General case}

The mode visibility is a way to express the mean value of the
squared amplitude of modes with a given degree compared to radial
modes. For a dipole mode, we define
\begin{equation}\label{eqt-def-visi-ampl}
    \Vdip = {A_1^2\over A_0^2}\ \Vgeo
    ,
\end{equation}
where $\Vgeo$ includes the contributions of several physical
effects: a geometrical factor that depends on the angular degree,
the limb-darkening, and the bolometric correction. $A_0$ and $A_1$
are the intrinsic amplitudes of the radial and dipole modes,
respectively. For stars exhibiting stochastically excited pure
acoustic modes, these amplitudes are supposed to be equal
\citep{2008A&A...478..163B}. In such a case, the visibility
$\Vdip$ reduces to $\Vgeo$.

A simple way to address the depressed modes consists in
normalizing the dipole visibility with respect to the nominal
expected value. We therefore use the same definition as
\cite{2015Sci...350..423F} for expressing the relative visibility
of depressed modes. In a first step, we consider the individual
visibilities of the mixed modes
\begin{equation}\label{eqt-ratio}
    \vm = {{\Vdip}\ind{depressed} \over
    \Vgeo}
    .
\end{equation}
where the mixed order $\nm$ labels the mixed mode \citep[Eqs.
(4.60)-(4.63) of][]{2015EAS....73....3M}. Strictly speaking, $\vm$
should be referred to as a ratio of the squared amplitudes of
dipole modes compared to radial modes, since the contribution of
the different visibility terms (aspect ratio of the spherical
harmonics and limb-darkening coefficients) is removed by the ratio
to $\Vgeo$. However, we keep the term visibility for the sake of
simplicity.

For red giants, \cite{2011A&A...531A.124B} computed $\Vgeo \simeq
1.54$, assuming that only acoustic modes are present. In evolved
stars, the situation is in fact complicated by the mixed modes.
The contribution of all mixed modes associated with a given
pressure radial order $\np$ can be expressed as
\begin{equation}\label{eqt-total}
   v = \sum_{\nm\in\{\dens\}} \vm
   .
\end{equation}
The sum is made in the ensemble $\{\dens\}$ of the $\dens$ mixed
modes associated with a given pressure radial order. They lie in
the $\Dnu$-wide frequency range between two radial modes: for the
pressure radial order around $\numax$, $\dens = \Dnu \Tg^{-1}
\numax^{-2}$, where $\Dnu$ is the mean large separation of
pressure modes, $\Tg$ is the period spacing of gravity modes, and
$\numax$ is the frequency of maximum oscillation signal.

The mixed-mode visibility $\vm$ introduced by
Eq.~(\ref{eqt-ratio}) has been investigated in previous work
\citep{2009A&A...506...57D,2014ApJ...781L..29B,2014A&A...572A..11G},
which shows
\begin{equation}\label{eqt-ratio-benomar}
   \vm \simeq
   \left(\frac{\Gamma_0}{\Gamma_\nm}\right)
   \left(\frac{I_0}{I_\nm}\right)^2
   ,
\end{equation}
where $\Gamma_0,\Gamma_\nm$ ($I_0,I_\nm$) are the linewidths
(inertia) of the radial and dipole modes, respectively. This
equation is valid whenever the dipole mode is resolved or not, but
assumes that the driving is the same for radial and dipole modes.
To go further, we have to examine two cases, depending on the
assumption on the dipole modes.

\subsection{Normal dipole mixed modes\label{cas-normal}}

Following \cite{2015A&A...579A..31B}, we may consider that the
work performed by the gas during one oscillation cycle, associated
with surface damping, is the same for all modes, so that
Eq.~(\ref{eqt-ratio-benomar}) is simplified into
\begin{equation}\label{eqt-ratio-normal}
   \vm \simeq {I_0 \over I_\nm} \simeq {\Gamma_\nm \over \Gamma_0}
   ,
\end{equation}
from which we retrieve that the individual visibilities of mixed
modes are small, with smaller mode widths and larger inertia than
radial modes.

From observations, \cite{2012A&A...537A..30M} have shown that the
contribution of all mixed modes associated with a given pressure
radial order $\np$ ensures $v=1$. Here, we can demonstrate this,
using the relation between inertia and the function $\zeta$ that
governs the mixed-mode spacings and the rotational splittings
\citep{2013A&A...549A..75G,2015A&A...580A..96D,2015A&A...584A..50M}.
From $I_0/I_\nm = 1-\zeta$, we have
\begin{equation}\label{eqt-equipartition0}
   v = \sum_{\nm\in\{\dens\}} (1-\zeta)
   .
\end{equation}
The $\dens$ mixed modes in  the ensemble $\{\dens\}$ between two
consecutive radial modes correspond to 1 pure pressure dipole mode
and $(\dens-1)$ pure gravity dipole modes. The period difference
between the two radial modes can be estimated in two ways, either
considering the sum $(\dens-1)\,\Tg$ for the pure gravity modes,
or considering the sum of the mixed-mode period spacings:
$\sum_{\nm\in\{\dens\}} \Delta P = (\sum_{\nm\in\{\dens\}}
\zeta)\; \Tg$, according to \cite{2015A&A...584A..50M}. Hence, we
get
\begin{equation}\label{eqt-equipartition}
   v = \sum_{\nm\in\{\dens\}} (1-\zeta)
   = \dens - (\dens -1 )
   = 1.
\end{equation}
This result proves that, despite the mixed nature of the dipole
modes, their total visibility matches the expected visibility of
the corresponding pure pressure mode. So, energy equipartition is
preserved for the normal mixed modes.

\subsection{A particular case: suppression of the oscillation in the core
\label{full_suppression}}

The possibility of the suppression of the oscillation in the core
was first investigated by \cite{1989nos..book.....U}. Their
equations (16.62)-(16.65) consider the effect of a wave leakage in
the core of an acoustic mode trapped in the convective envelope.
The limit of oscillation suppression in the core  implies that
only pressure dipole modes are present since mixed modes are
necessarily cancelled out. In that case, $I_1 \simeq I_0$, so that
Eq.~(\ref{eqt-ratio-benomar}) rewrites
\begin{equation}\label{eqt-ratio-suppression}
   v = {\Gamma_0 \over \Gamma_1}
   .
\end{equation}
The damping in the core, whatever it is, can be written
\begin{equation}\label{cas_fuller}
   v = \frac{\Gamma_0}{\Gamma_1^{\rm env}+\Gamma_1^{\rm core}}
        \simeq \frac{\Gamma_0}{\Gamma_0+\Gamma_1^{\rm core}}
           ,
\end{equation}
where  $\Gamma_1^{\rm env}\simeq \Gamma_0$ and $\Gamma_1^{\rm
core}$ are the damping contributions in the envelope and in the
core, respectively.

\subsubsection{Damping and transmission}

Equation~(\ref{cas_fuller}) can be rewritten
    \begin{equation}\label{cas_fuller2}
    v =  \frac{1}{1+x} \quad {\rm with } \quad x = \frac{\Gamma_{\rm 1}^{\rm core}}{\Gamma_0}
    .
    \end{equation}
When all energy transmitted in the core is absorbed or damped,
following \cite{1989nos..book.....U} we get
\begin{equation}\label{ratio_dampings}
    x
    =
    \omega \tau\ind{a} \left(\int_\cavp k_r \,\diff r \right)^{-1}
    \frac{1}{4} \exp \left(-2\int_\eva \kappa \,\diff r \right)
    ,
\end{equation}
where $\cavp$ denotes integration in the acoustic cavity and
$\eva$ in the evanescent region,  $k_r$ is the radial wave number,
$\kappa^2 = -k_r^2$, $\omega$ is the mode frequency, and
$\tau\ind{a}$ is the e-folding damping time of the radial mode
amplitude. At first order, the first integral of
Eq.~(\ref{ratio_dampings}) equals $\omega / 2\Delta \nu$, where
$\Delta \nu$ is the large separation. Thus,
Eq.~(\ref{ratio_dampings}) becomes
\begin{equation}
   x
   =
   \frac{\Delta \nu \, \tau\ind{a}}{2} \exp \left(-2\int_\eva \kappa {\rm d}r \right)
   .
\end{equation}
We then introduce the e-folding damping time of the mode energy,
$\tau_0 = \tau\ind{a}/2$ and get
\begin{equation}\label{eq-visibility}
    v =  \left(1+ \Delta \nu \, \tau_0\, T^2\right)^{-1}
    ,
\end{equation}
where the transmission in the evanescent region is defined by
\begin{equation}
  T = \exp \left(-\int_\eva \kappa {\rm d}r \right)
  .
\end{equation}
Equation~(\ref{eq-visibility}), derived from the relative
visibility of depressed dipole modes with respect to radial modes,
is similar to Eq.~(2) of \cite{2015Sci...350..423F}, derived from
the ratio between the depressed and normal dipole modes.

\begin{figure}
\includegraphics[width=8.8cm]{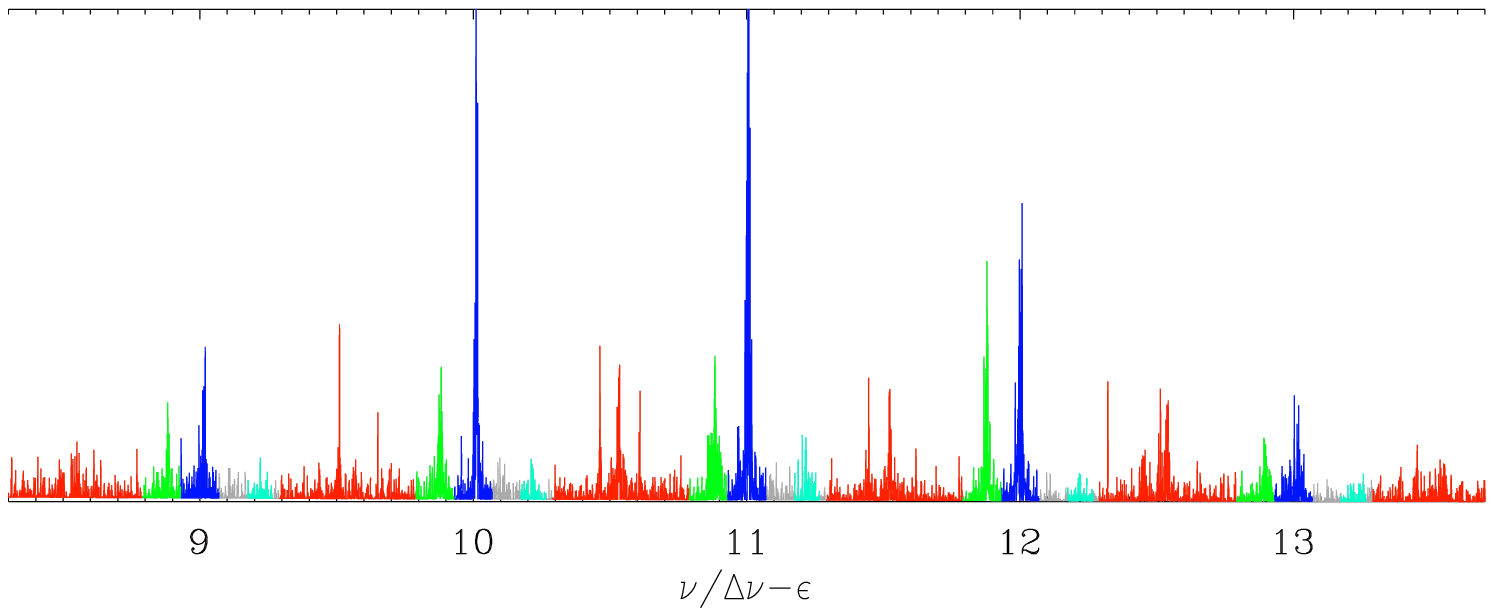}
\includegraphics[width=8.8cm]{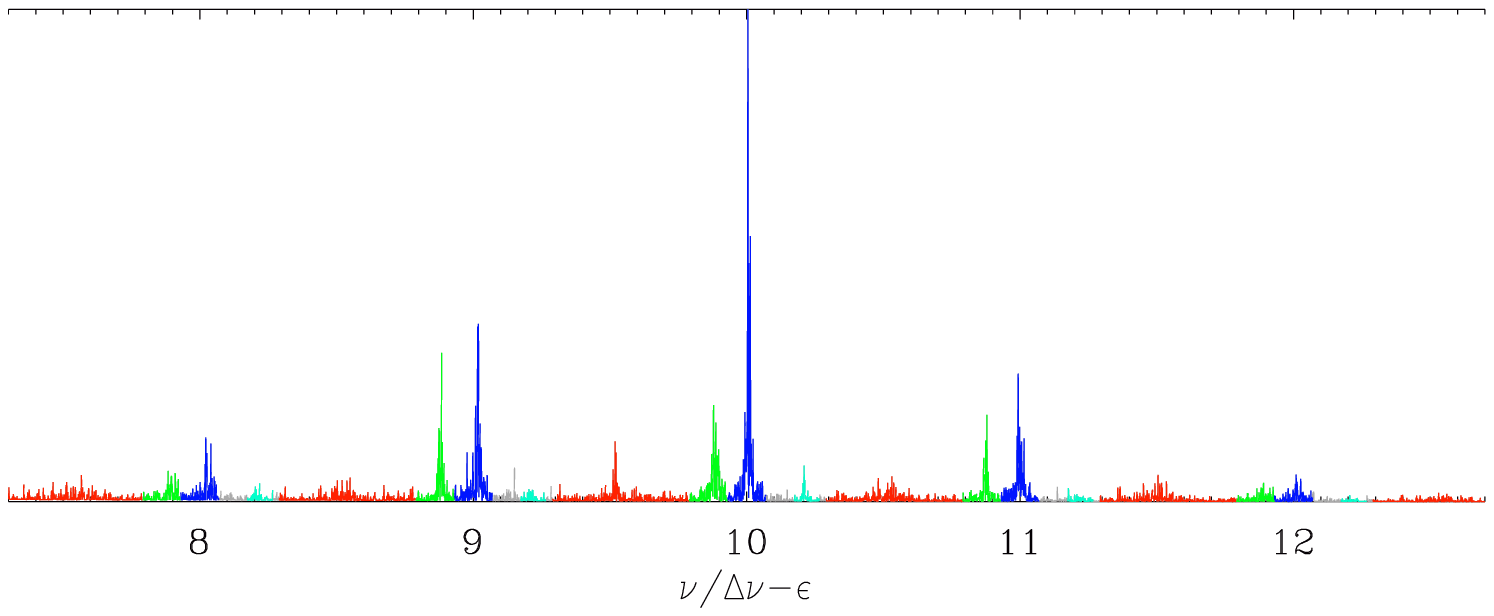}
\includegraphics[width=8.8cm]{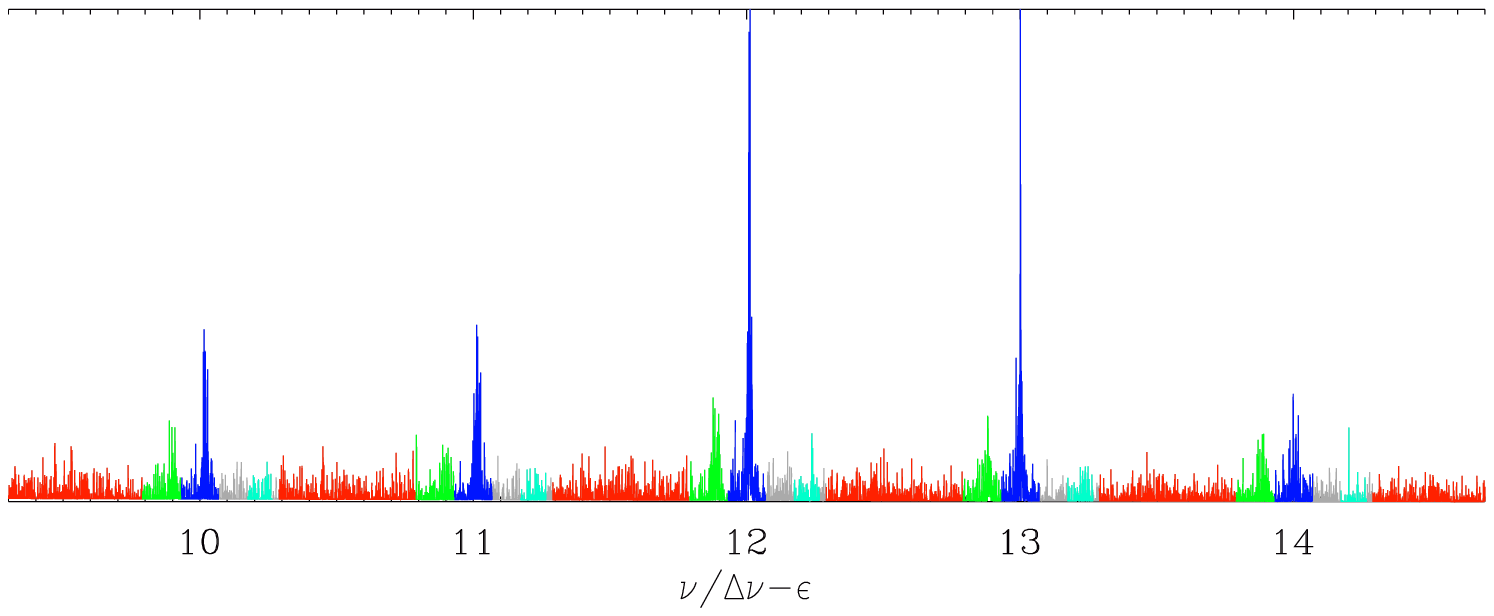}
\caption{Examples of RGB stars with low dipole-mode amplitudes.
Spectra are plotted as a function of the reduced frequency
$\nu/\Dnu-\varepsilon$, so that radial modes (shown in blue) are
close to integer values; quadrupole modes are plotted in green,
octupole modes in cyan; mixed modes can be found everywhere but
their contribution is measured in the frequency range plotted in
red only. The mean background component was subtracted.
{\sl top)} the mixed mode pattern of KIC 9279486  can be fitted in
order to extract seismic global parameters;
{\sl middle)} mixed modes in KIC 6026983 are evidenced but their
pattern cannot be fit;
{\sl bottom)} dipole modes of KIC 5810513 have very small
amplitudes but can be unambiguously detected and identified as
mixed modes with the method discussed in Section
\ref{hidden}.}\label{fig-type}
\end{figure}

\begin{figure}
\includegraphics[width=8.8cm]{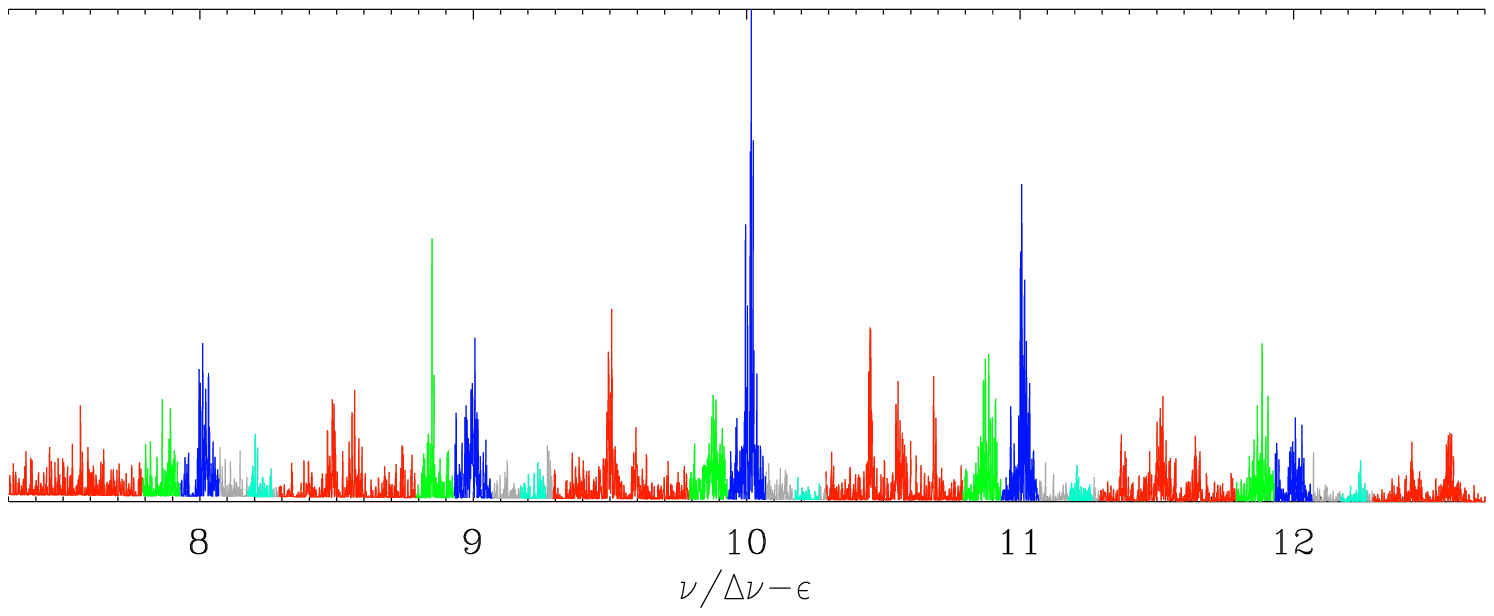}
\includegraphics[width=8.8cm]{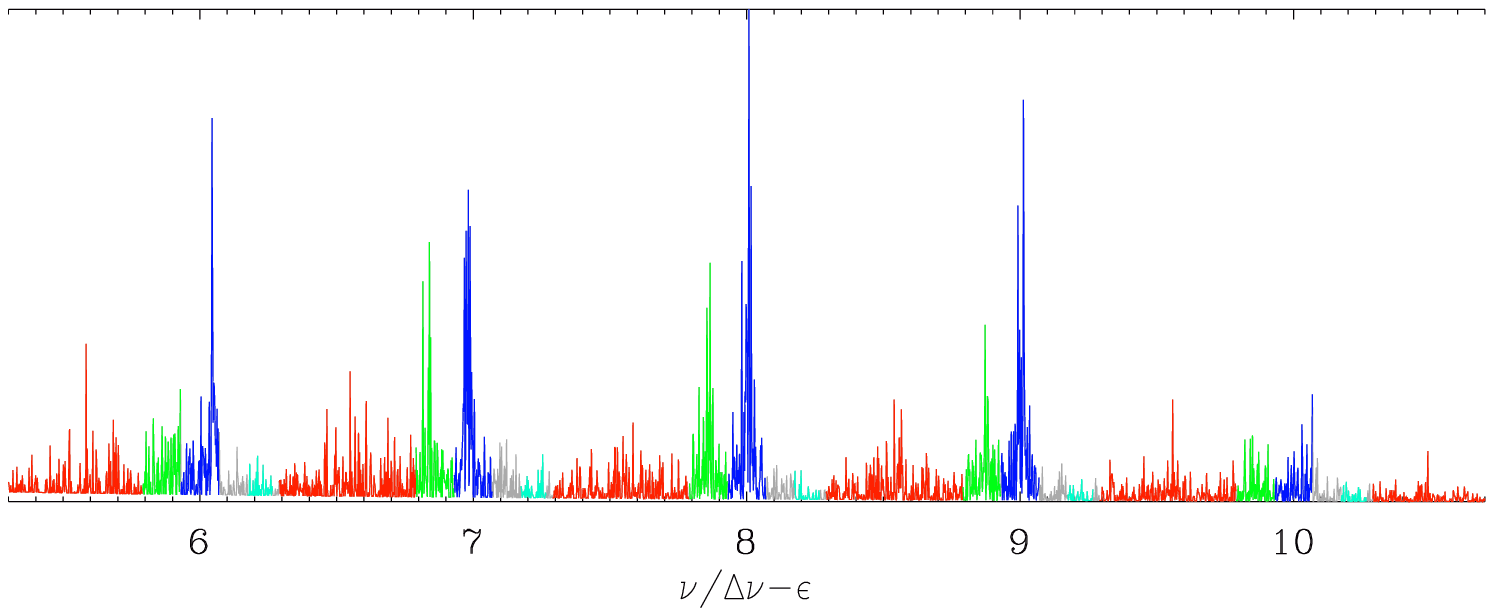}
\caption{Same as Fig.~\ref{fig-type} but for core-helium burning
stars.
{\sl top)} the mixed mode pattern of the secondary-clump star KIC
8522050 can be fitted in order to extract seismic global
parameters;
{\sl bottom)} depressed mixed modes in the clump star KIC 2693261
are apparent but their pattern cannot be fit.\label{fig-type2}}
\end{figure}

\subsubsection{Link with observable parameters}

We can match the value of $T^2$ with the coupling factor $q$ of
mixed modes \citep{1989nos..book.....U}:
\begin{equation}\label{eqt-relation-q-T0}
    T^2 = 4 q
    .
\end{equation}
The visibility can thus be expressed as a function of the seismic
observables $q$ and $\Gamma_0$
\begin{equation}\label{eq-visibility-low-q}
    v = \left(1 + 2 q\ {\Dnu \over \pi \Gamma_0}\right)^{-1}
    .
\end{equation}
The coupling factor $q$ is obtained from the asymptotic expansion
of mixed modes
\begin{equation}\label{eqt-asymp}
    \tan\theta\ind{p} = q \tan\theta\ind{g}
    ,
\end{equation}
where the phases $\theta\ind{p}$ and $\theta\ind{g}$ refer,
respectively, to the pressure- and gravity-wave contributions
\citep{2015A&A...584A..50M}. The radial mode width $\Gamma_0$,
measured as the full width at half maximum in the power density
spectrum, is related to the radial mode lifetime by
\citep[see][]{2015EAS....73..111S}
\begin{equation}\label{eq-gamma0-tau0}
  \Gamma_0 = {1 \over 2 \pi\tau_0}
  .
\end{equation}
In fact, Eqs. (\ref{eqt-relation-q-T0}) and
(\ref{eq-visibility-low-q}) are no longer valid when the extent of
the evanescent region is limited, so that strong coupling occurs.
In that case \citep[see][]{takata},
\begin{equation}\label{eqt-relation-T-q}
    T^2 = {4 q\over (1+q)^2}
    .
\end{equation}
Following Eq.~(71) of \cite{takatab}, we also have to replace
$T^2$ by $-\ln (1 - T^2)$ in case of full suppression of the
oscillation in the core. So, Eq.~(\ref{eq-visibility}) becomes
\begin{equation}\label{eq-visibility-q}
    v = \left(1 - \ln\left[1- {4 \,q\over (1+q)^2}\right] { \Dnu \over 2 \pi \Gamma_0}\right)^{-1}
    .
\end{equation}
Contrary to Eq.~(\ref{eq-visibility}), this expression ensures a
null visibility in case of total transmission  ($T=q=1$).

\begin{figure*}
\includegraphics[width=8.8cm]{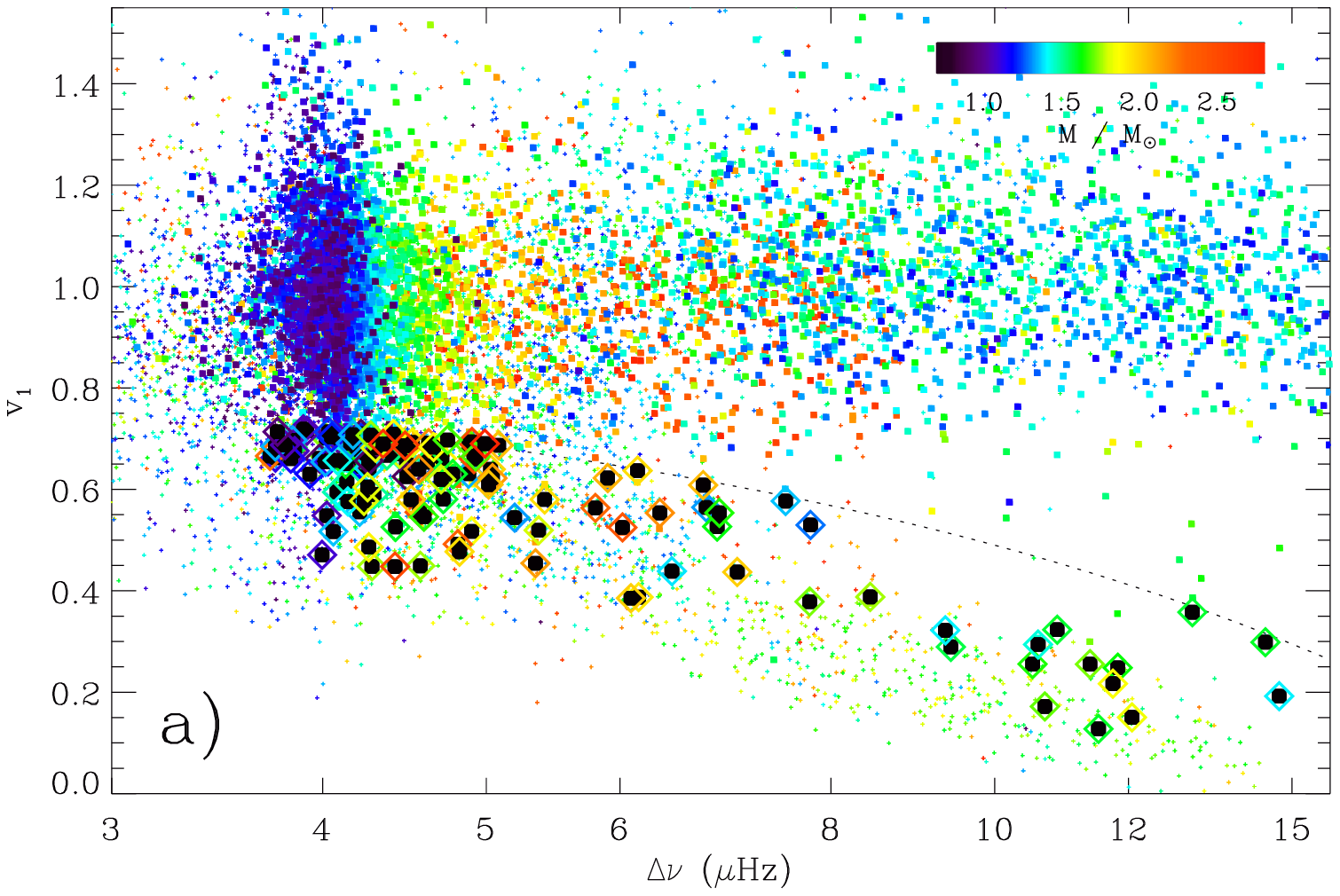}
\hfill
\includegraphics[width=8.8cm]{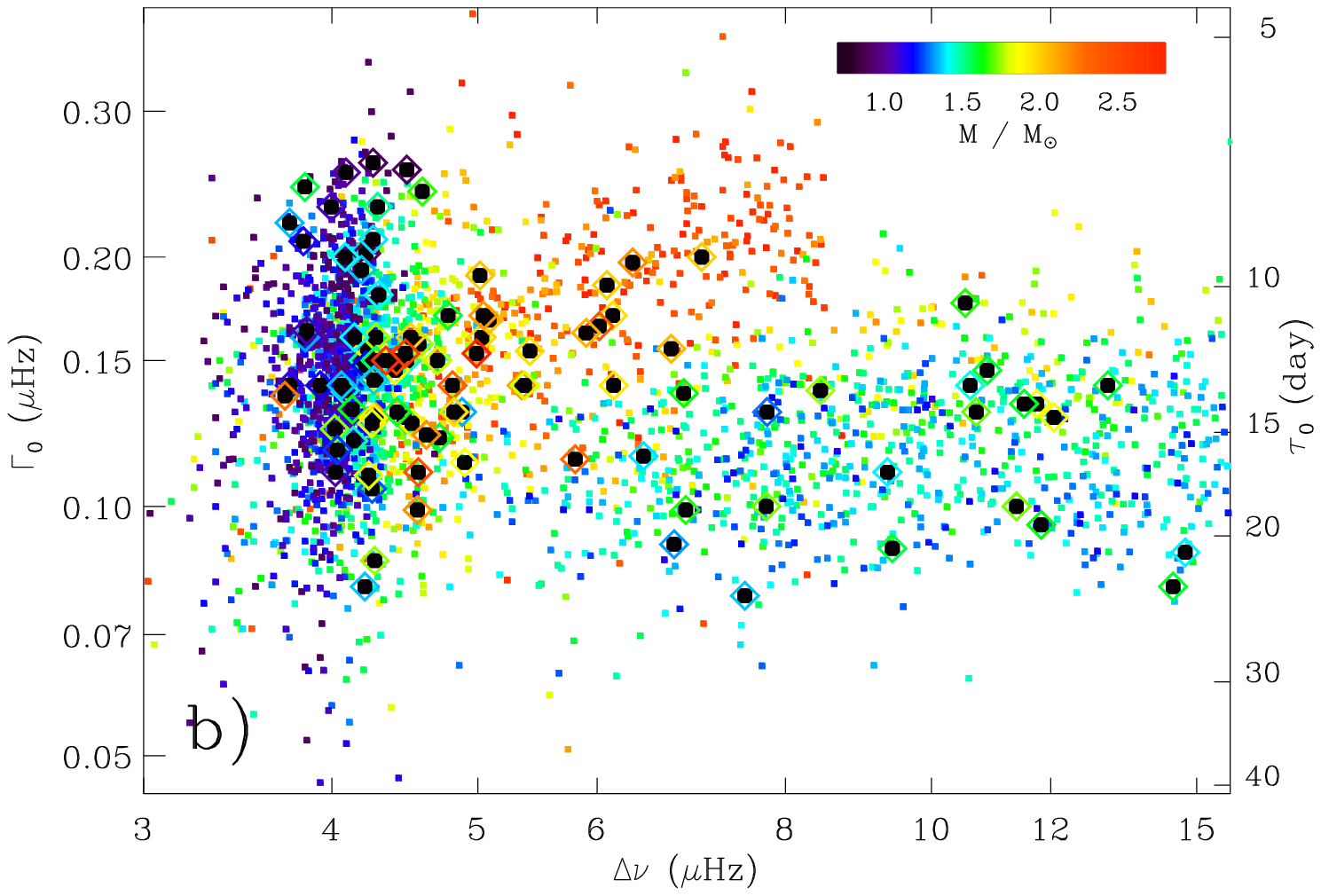}
\smallskip
\includegraphics[width=8.8cm]{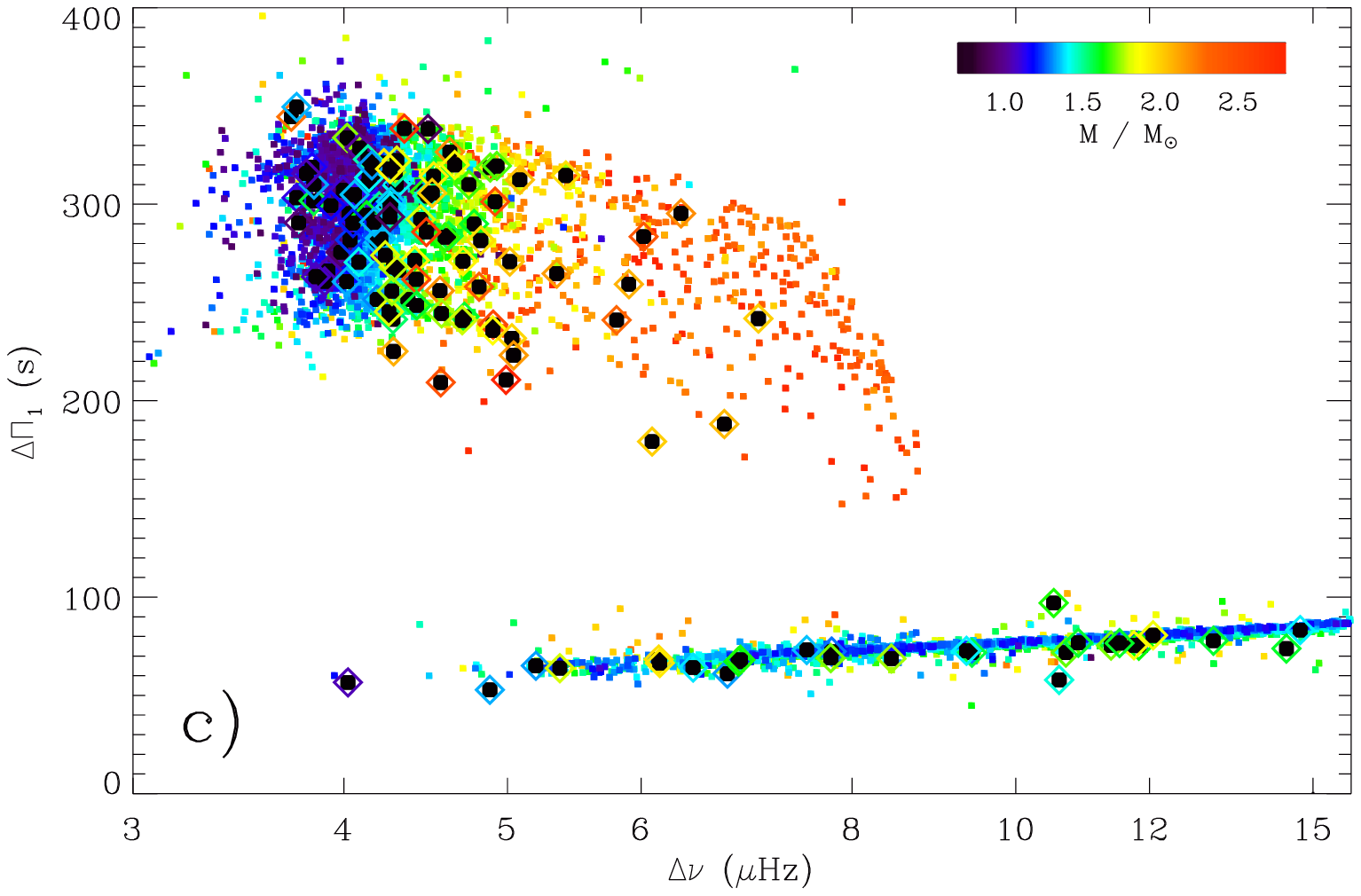}\quad
\hfill
\includegraphics[width=8.8cm]{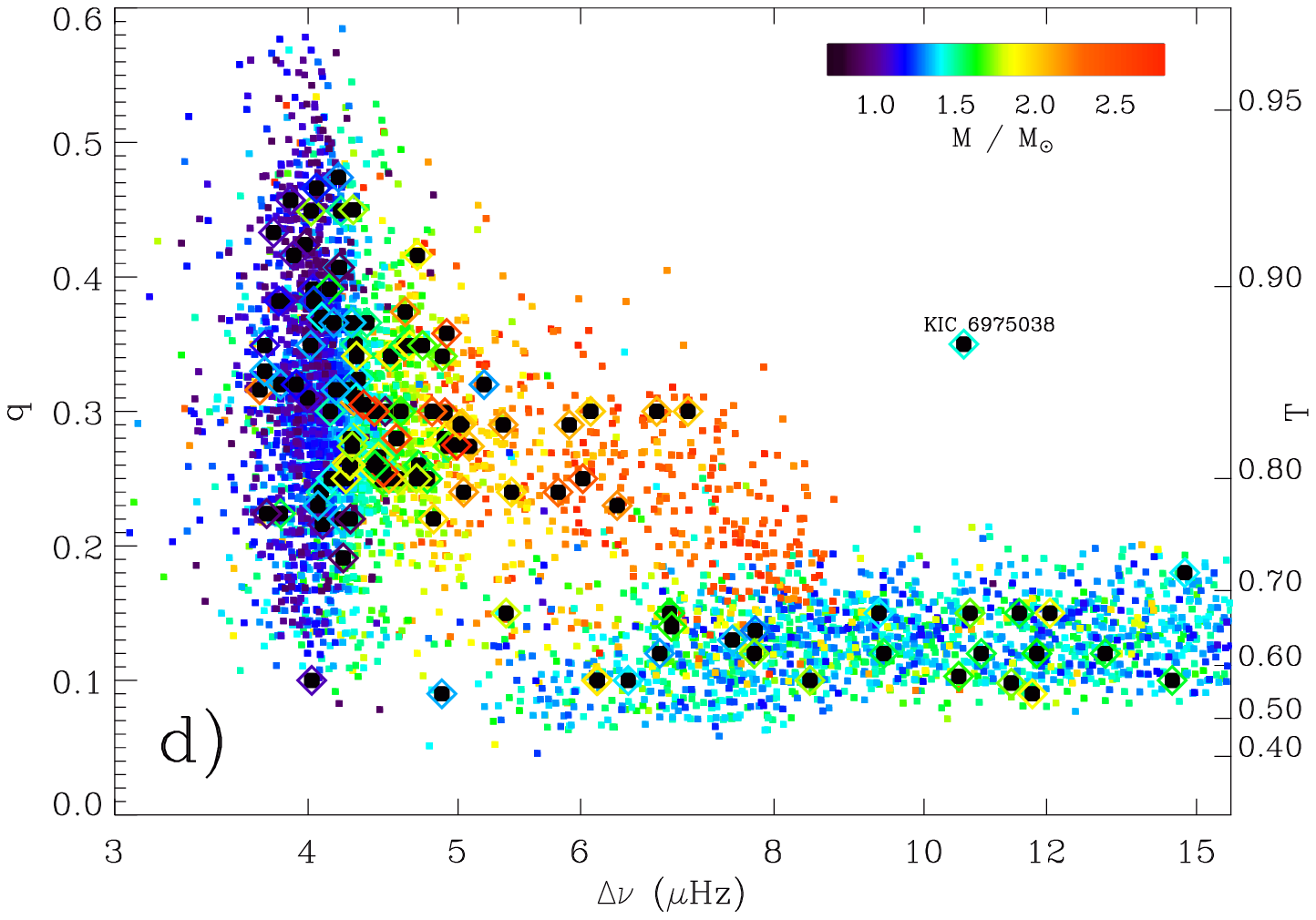}
\caption{\emph{a)} Dipole mode visibility $v_1$ as a function of
the large separation $\Dnu$. The color codes the mass, determined
with seismic scaling relations. Small symbols represent \Kepler\
stars of the public data set; larger symbols represent stars of
the data set studied by \protect\cite{2016A&A...588A..87V}.
Diamonds with a dark center are stars showing depressed mixed
modes. The dashed line represents the limit defining low
visibilities;
\emph{b)} Radial mode width $\Gamma_0$ as a function of the large
separation $\Dnu$ of the data set studied by
\protect\cite{2016A&A...588A..87V}. The right y-axis provides the
radial mode lifetime;
\emph{c)} Same as \emph{b)} for the period spacings $\Tg$;
\emph{d)} Same as \emph{b)} for the coupling factor $q$. The
outlying value of $q$ for KIC 6975038 is discussed in Section
\ref{section-visibility}.  \label{fig-deprime}}
\end{figure*}

\section{Seismic observables\label{observations}}

Previous observations have reported that the depressed modes in
the red giant KIC 8561221 are mixed \citep{2014A&A...563A..84G}.
Such observations question the hypothesis of oscillation
suppression: if the low visibility derives from the suppression of
the oscillation in the radiative core, mixed modes cannot be
established. Therefore, we first aim at identifying the prevalence
of red giants with depressed mixed modes. Then, we also use
different observations to assess the properties of depressed
dipole modes, in order to determine whether they are mixed or not.
Figure \ref{fig-type} illustrates the different types of stars we
intend to work with: either with depressed mixed modes that can be
identified, or with depressed mixed modes that cannot be fitted,
or without clear evidence of mixed modes. Figure \ref{fig-type2}
provides examples for core-helium burning stars.

\subsection{Identification of low visibilities}

The first step for the search of stars with dipole mixed modes
consists in the measurement of reduced visibilities, as defined by
Eq.~(\ref{eqt-total}), with the method of
\cite{2012A&A...537A..30M}. In short, squared amplitudes are
estimated from the integration of the power spectrum density over
the frequency range covering the different modes, after
subtraction of background. We obtained the total dipole
visibilities $\Vdip$ for about 12\,500 red giants of the \Kepler\
public data (Fig.~\ref{fig-deprime}a), from which we could
identify the population of  stars on the red giant branch (RGB)
with normal visibilities and the family of stars with low
visibility, in agreement with \cite{2012A&A...537A..30M} and
\cite{2016PASA...33...11S}. Red giants with normal amplitudes have
a total dipole visibility close to $\Vgeo \simeq 1.54$, with a
very small dependence in $\Teff$, $\log g$ and $Z$. Using
effective temperatures of \cite{2014ApJS..211....2H}, we found
that the normal (not depressed) visibilities of red giants,
integrated for one pressure radial order, follow the mean trend
\begin{equation}\label{eqt-visi-trend}
    \langle \Vgeo \rangle
    \simeq 1.54 -  { \Teff- 4850 \over 4100}
    ,
\end{equation}
where the brackets indicate the average value derived from a
linear fit with the effective temperature $\Teff$ expressed in
kelvin. We note that this result is very close to the predictions
of \cite{2011A&A...531A.124B}.

The mean value of the normal visibility was then used to derive
reduced integrated observed visibilities
\begin{equation}\label{eqt-visi-def}
    v_1 = {\overline{\sum \Vdip} \over \langle \Vgeo\rangle}
    ,
\end{equation}
where, for the numerator, the sum matches all dipole modes
associated to a given pressure radial order and the overbar
represents the mean value for the different radial orders where
dipole modes are observed. Now, $v_1$ can be compared to $v$
(Eq.~\ref{eqt-total}). As stated by previous work, the limit
between normal and low visibility is clear on the RGB, despite the
presence of a few stars lying in the no man's land between normal
and reduced visibilities; in the red clump, we chose to define
low-visibility stars by $v_1 \le 0.85 - 0.04 \,\Dnu $ (with $\Dnu$
expressed in $\mu$Hz). We tested that changing the threshold value
does not significantly change the conclusions of the work.

\begin{table*}[t]
\caption{Seismic properties of \nfit\ stars showing depressed
mixed modes}\label{tab-properties} \scriptsize %
\begin{tabular}{rcrrrrrrrrrr}
\hline
 KIC & evolutionary& $\numax$ & $\Dnu$ & $\Tg$& $q$ & $M$ & $\Gamma_0$ & $\dnurot$ & $v_1$ & $v$ & $v$ \\
     & stage        & ($\mu$Hz)&($\mu$Hz)& (s) & & ($M_\odot$) & ($\mu$Hz)  & (nHz)& & low & strong \\
     & (1)          &          &         &     & (2)&  & (3) & (4) \\
 \hline
  2573092& RC&   35.3&   4.08&  293.8&   0.24&   1.48&   0.14&   35&   0.60&  0.181&  0.178\\
  2858129&RC2&   40.5&   4.22&  320.9&   0.26&   1.93&   0.16&   80&   0.58&  0.181&  0.178\\
  2992350& RC&   45.9&   4.72&  242.8&   0.26&   1.59&   0.12&   35&   0.58&  0.134&  0.132\\
  3443483&RGB&  132.2&  10.71&   71.8&   0.15&   1.71&   0.13&  230&   0.16&  0.113&  0.112\\
  3532734&RGB&  145.2&  11.83&   75.3&   0.12&   1.63&   0.09&    0&   0.23&  0.095&  0.095\\
  3660245& RC&   33.3&   4.22&  297.0&   0.25&   1.19&   0.15&   35&   0.65&  0.181&  0.178\\
  3758505&RGB&  157.4&  12.06&   80.6&   0.15&   1.84&   0.13&  500&   0.14&  0.100&  0.099\\
  4180939&RC2&   65.2&   5.80&  241.0&   0.24&   2.29&   0.11&   50&   0.55&  0.114&  0.112\\
  4587050&RGB&  191.2&  14.47&   73.8&   0.10&   1.60&   0.08&  530&   0.26&  0.080&  0.080\\
  4761301& RC&   30.8&   4.49&  338.3&   0.30&   0.76&   0.25&   65&   0.63&  0.229&  0.224\\
  4772722& RC&   37.3&   4.30&  310.0&   0.31&   1.42&   0.18&   38&   0.67&  0.175&  0.170\\
  5007332&RGB&   96.9&   8.44&   68.7&   0.10&   1.76&   0.14&  310&   0.37&  0.204&  0.204\\
  5179471&RC2&   47.0&   4.56&  256.0&   0.28&   2.17&   0.10&  110&   0.64&  0.109&  0.106\\
  5295898&RGB&  142.9&  11.39&   75.3&   0.10&   1.72&   0.10&  210&   0.23&  0.123&  0.123\\
  5306667& RC&   39.5&   4.28&  268.0&   0.28&   1.77&   0.14&   25&   0.67&  0.158&  0.154\\
  5339823& RC&   40.4&   4.28&  317.2&   0.45&   1.76&   0.16&   80&   0.45&  0.115&  0.108\\
  5620720&RC2&   57.0&   5.35&  264.7&   0.29&   2.14&   0.14&   45&   0.45&  0.124&  0.121\\
  5881079&RGB&   67.2&   6.14&   68.3&   0.10&   1.96&   0.17&    0&   0.62&  0.303&  0.302\\
  5949964& RC&   37.8&   4.29&  316.3&   0.35&   1.50&   0.23&   40&   0.59&  0.194&  0.187\\
  6037858& RC&   43.3&   4.57&  244.4&   0.25&   1.76&   0.16&   50&   0.45&  0.178&  0.174\\
  6130770& RC&   35.1&   4.13&  322.9&   0.30&   1.48&   0.12&   40&   0.62&  0.132&  0.128\\
  6210264& RC&   40.7&   4.78&  290.0&   0.25&   1.60&   0.17&   60&   0.63&  0.183&  0.180\\
  6232858&RC2&   50.0&   4.90&  235.8&   0.28&   1.87&   0.11&   55&   0.51&  0.115&  0.112\\
  6610354&RC2&   46.9&   4.57&  209.3&   0.28&   2.44&   0.11&   65&   0.55&  0.119&  0.116\\
  6975038&RGB&  128.4&  10.61&   57.9&   0.35&   1.44&   0.14&  280&   0.27&  0.056&  0.054\\
  7512378& RC&   32.5&   3.84&  310.0&   0.32&   1.34&   0.16&   45&   0.70&  0.170&  0.165\\
  7515137&RGB&   67.2&   6.75&   61.1&   0.12&   1.32&   0.09&    0&   0.55&  0.149&  0.148\\
  7693833&RGB&   31.8&   4.02&   56.6&   0.10&   1.05&   0.11&    0&   0.55&  0.301&  0.300\\
  7746983&RGB&  188.6&  14.74&   83.3&   0.18&   1.40&   0.09&  190&   0.17&  0.050&  0.049\\
  8009582& RC&   35.2&   4.21&  282.4&   0.22&   1.36&   0.08&   30&   0.56&  0.120&  0.118\\
  8025383& RC&   36.1&   4.14&  313.5&   0.25&   1.41&   0.16&   25&   0.58&  0.195&  0.192\\
  8283646&RGB&   67.2&   6.15&   66.5&   0.10&   1.93&   0.14&  100&   0.38&  0.263&  0.263\\
  8391175&RGB&   87.4&   7.77&   69.0&   0.12&   1.72&   0.10&    0&   0.36&  0.144&  0.144\\
  8396782&RC2&   82.3&   7.04&  241.8&   0.30&   2.00&   0.20&  200&   0.42&  0.129&  0.126\\
  8432219& RC&   42.3&   4.59&  283.2&   0.30&   1.66&   0.24&   45&   0.55&  0.215&  0.210\\
  8476202&RGB&  109.7&   9.42&   71.7&   0.12&   1.57&   0.09&    0&   0.27&  0.110&  0.110\\
  8522050&RC2&   75.5&   6.72&  188.1&   0.30&   2.06&   0.16&   60&   0.59&  0.108&  0.105\\
  8564277& RC&   31.9&   4.00&  307.0&   0.31&   1.09&   0.23&   25&   0.47&  0.226&  0.220\\
  8564559& RC&   45.8&   4.70&  240.9&   0.25&   1.79&   0.15&   40&   0.62&  0.167&  0.164\\
  8636174&RC2&   43.9&   4.51&  305.7&   0.25&   2.00&   0.16&   35&   0.58&  0.182&  0.179\\
  8687248&RGB&  170.0&  13.10&   77.8&   0.12&   1.59&   0.14&    0&   0.32&  0.123&  0.122\\
  8689599& RC&   30.4&   3.93&  299.3&   0.32&   1.17&   0.14&   35&   0.63&  0.149&  0.144\\
  8771414& RC&   38.6&   4.23&  274.0&   0.25&   1.82&   0.11&   28&   0.58&  0.139&  0.137\\
  8827934&RGB&   55.0&   5.37&   63.8&   0.15&   1.80&   0.14&    0&   0.51&  0.214&  0.213\\
  9115334&RC2&   67.5&   6.09&  179.2&   0.30&   2.00&   0.19&  220&   0.38&  0.137&  0.134\\
  9176207&RC2&   59.2&   5.41&  314.6&   0.24&   1.96&   0.15&  100&   0.57&  0.157&  0.154\\
  9229592&RGB&   72.0&   6.85&   67.2&   0.15&   1.62&   0.14&  230&   0.51&  0.173&  0.172\\
  9279486&RGB&  132.4&  10.89&   76.8&   0.12&   1.59&   0.15&  200&   0.30&  0.149&  0.149\\
  9291830&RC2&   47.5&   4.41&  261.8&   0.30&   2.50&   0.15&   70&   0.44&  0.151&  0.147\\
  9581849& RC&   34.6&   4.08&  270.5&   0.37&   1.43&   0.20&   30&   0.66&  0.172&  0.165\\
  9650046&RC2&   68.2&   6.02&  283.5&   0.25&   2.42&   0.17&   90&   0.51&  0.147&  0.144\\
  9711269&RGB&   64.1&   6.44&   64.2&   0.10&   1.39&   0.12&    0&   0.43&  0.219&  0.218\\
  9719858&RC2&   47.8&   4.48&  285.8&   0.25&   2.62&   0.15&   90&   0.68&  0.174&  0.171\\
  9947511& RC&   30.7&   3.75&  349.5&   0.33&   1.35&   0.22&   30&   0.69&  0.218&  0.212\\
 10029821&RC2&   64.6&   5.90&  259.3&   0.29&   2.08&   0.16&   50&   0.61&  0.129&  0.126\\
 10091729&RC2&   72.8&   6.34&  295.3&   0.23&   2.18&   0.20&   60&   0.54&  0.175&  0.173\\
 10420655& RC&   37.5&   4.26&  298.6&   0.28&   1.39&   0.21&   50&   0.64&  0.217&  0.212\\
 10422589&RC2&   50.2&   5.04&  223.1&   0.24&   2.15&   0.17&   60&   0.62&  0.181&  0.178\\
 10469976&RC2&   52.7&   4.81&  258.0&   0.30&   2.35&   0.14&   30&   0.49&  0.132&  0.129\\
 10528917&RGB&   76.4&   7.52&   73.1&   0.13&   1.36&   0.08&  220&   0.56&  0.111&  0.111\\
 10653383& RC&   40.1&   4.42&  248.5&   0.26&   1.60&   0.13&   30&   0.53&  0.151&  0.148\\
 10854564& RC&   29.9&   4.26&  293.5&   0.22&   0.69&   0.26&   30&   0.66&  0.304&  0.300\\
 11413158&RC2&   58.1&   4.99&  210.6&   0.28&   2.86&   0.15&   80&   0.68&  0.149&  0.146\\
 11462972&RC2&   29.9&   4.26&  318.0&   0.26&   1.89&   0.13&   30&   0.49&  0.154&  0.151\\
 11519450&RGB&   72.8&   6.87&   68.1&   0.14&   1.65&   0.10&  180&   0.54&  0.139&  0.138\\
 11598312&RC2&   48.0&   4.82&  281.5&   0.22&   1.96&   0.13&   50&   0.47&  0.161&  0.159\\
 12058556&RGB&  105.0&   9.35&   72.7&   0.15&   1.40&   0.11&    0&   0.31&  0.110&  0.109\\
 12070510& RC&   35.3&   4.06&  304.9&   0.23&   1.36&   0.14&   50&   0.52&  0.191&  0.188\\
 12109388& RC&   40.9&   4.25&  245.3&   0.26&   1.82&   0.13&   50&   0.60&  0.152&  0.149\\
 12453551&RC2&   51.2&   5.02&  270.8&   0.29&   2.04&   0.19&   70&   0.60&  0.170&  0.166\\
 12691734&RC2&   46.8&   4.34&  338.5&   0.31&   2.73&   0.15&   50&   0.69&  0.151&  0.147\\
 \hline
\end{tabular}

- (1) RGB: red giant branch; RC: red clump; RC2: secondary red
clump.

- (2) Uncertainties on $q$ are of about $\pm$0.027 for stars on
the RGB and $\pm$0.057 in the red clump.

- (3) Uncertainties on $\Gamma_0$ are of about 30\,\%.

- (4) A null value for $\dnurot$ indicates that the rotational
splitting could not be measured because the star is seen pole-on.

\end{table*}

\begin{figure}
\includegraphics[width=8.8cm]{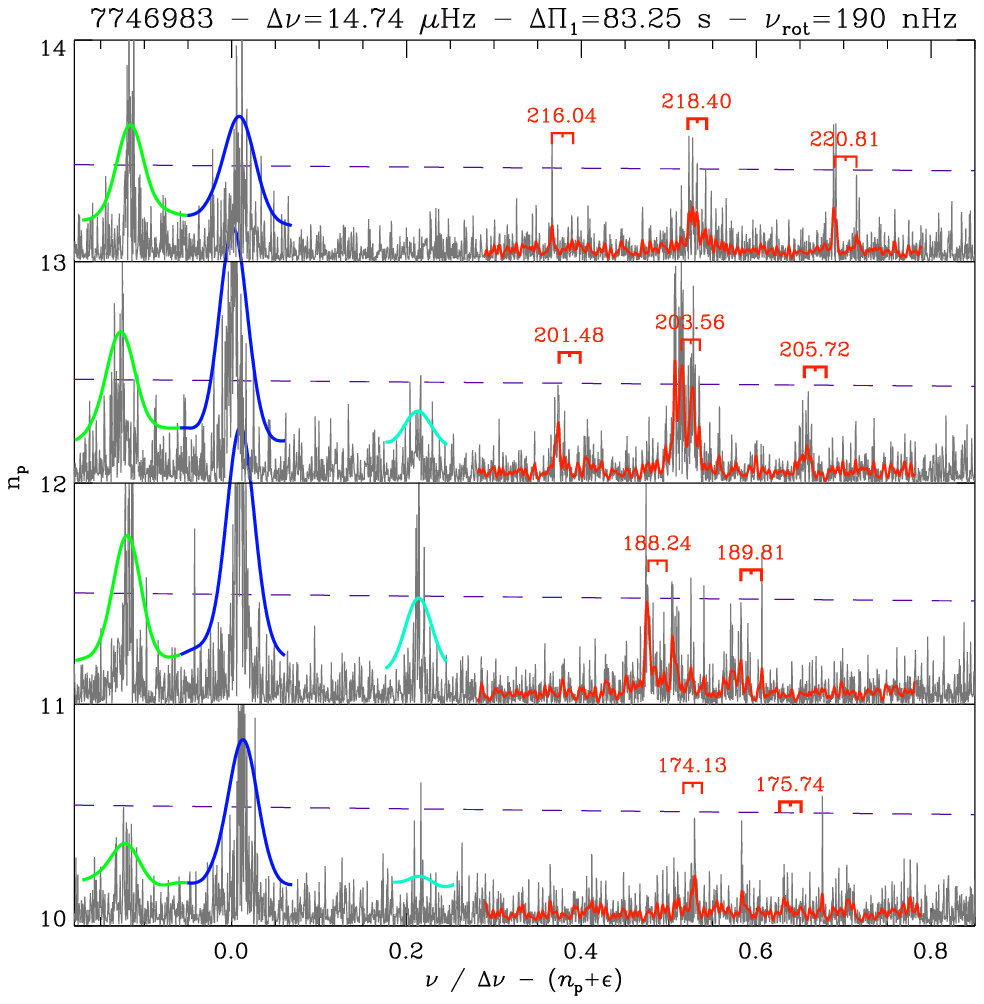}
\includegraphics[width=8.8cm]{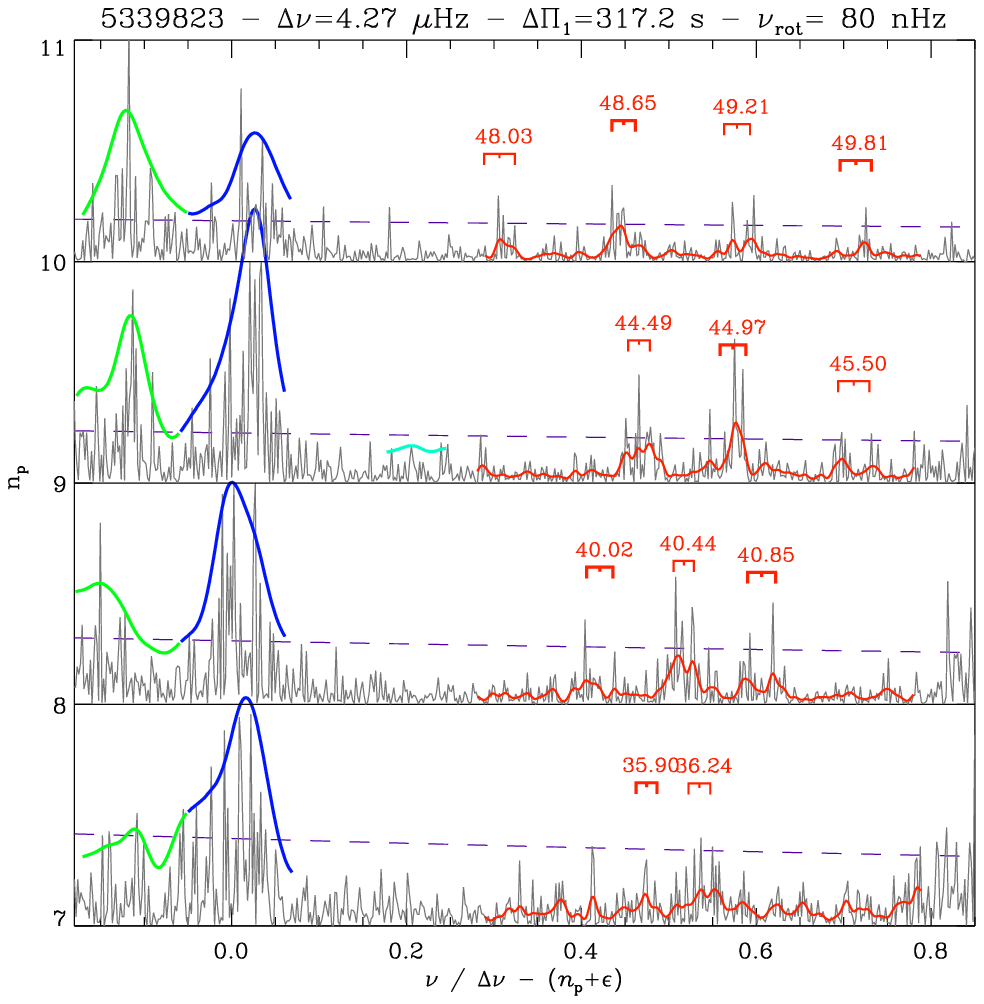}
\caption{Examples of complete fits of the asymptotic mixed-mode
pattern of red giants with low dipole-mode amplitudes. \'Echelle
diagrams are plotted as a function of the reduced frequency
$\nu/\Dnu-(\np+\varepsilon)$, so that radial modes are close to
integer values. The smoothed profile of mixed modes are plotted in
red, quadrupole modes in green, octupole modes in cyan. The dashed
gray line corresponds to 8 time the granulation background. Dipole
triplets are identified by the asymptotic cyclic frequency, in
$\mu$Hz, of the $m=0$ component.
{\sl top)} RGB star, KIC 7746983;
{\sl bottom)} Clump star, KIC 5339823.
}\label{fig-fits}
\end{figure}

\subsection{Fit of the mixed-mode pattern\label{depressed-stars}}

The systematic search for stars with depressed mixed modes was
derived from the recent work of \cite{2016A&A...588A..87V}, who
have measured the asymptotic period spacing of mixed modes for
about 6\,100 red giants. We then fitted the asymptotic mixed-mode
pattern in stars with reduced dipole visibilities.

The fit of mixed mode frequencies in red giants is usually made
easy by the use of the asymptotic expansion
\citep{1989nos..book.....U,2012A&A...548A..10M}, but is more
difficult in stars with depressed mixed modes because of the lower
signal-to-noise ratio induced by the low visibilities. However, we
managed to optimize this fitting process to obtain complete sets
of seismic parameters, including rotational splittings $\dnurot$
\citep{2012A&A...548A..10M}. Recent methods and results based on
four years of \Kepler\ observation were used to update previous
measurements
\citep{2014A&A...572L...5M,2015A&A...584A..50M,2016A&A...588A..87V}.
Stellar masses were estimated from the seismic scaling relation
with the method of \cite{2013A&A...550A.126M} in order to have a
better calibration than the solar calibration and to lower the
non-negligible noise induced by pressure glitches
\citep{2015A&A...579A..84V}. Scaling relations have been used with
the effective temperature of \cite{2014ApJS..211....2H}.

We were able to fit the asymptotic mixed-mode pattern, including
the rotational splittings, for \nfit\ red giants (Table
\ref{tab-properties}, Fig.~\ref{fig-fits}). This number represents
a small fraction of the 1109 stars with low amplitudes since
fitting all parameters of the asymptotic mixed-mode pattern is
highly demanding when amplitudes are depressed. Unsurprisingly,
owing to the aforementioned observational bias, our data set with
mixed modes on the RGB is biased toward high visibilities
(Fig.~\ref{fig-deprime}a). Conversely, as low visibilities of
clump stars are not as low as on the RGB, fitting their mixed
modes is easier. With this analysis, we can establish the
properties of stars with depressed mixed modes.

\paragraph{Mass}

Stars in our data set with depressed mixed modes present larger
masses than the typical mass distribution of CoRoT or \Kepler\ red
giants showing solar-like oscillations. Their median mass is
1.6\,$M_\odot$, above the median mass of the red giants observed
with \Kepler\ (1.4\,$M_\odot$). This agrees with the mass
distribution found by \cite{2016Natur.529..364S} for stars with
low-amplitude dipole modes.

\paragraph{Evolutionary stage}

\cite{2012A&A...537A..30M} have reported the identification of low
visibilities for stars on the RGB. Here, we report that stars with
depressed mixed modes appear at any evolutionary stages. We
identified depressed modes in secondary-clump stars, located on
the same low-visibility branch as RGB stars
(Fig.~\ref{fig-deprime}a). Due to the mass dependence of the
low-visibility stars, depressed modes are in fact over-represented
in the secondary red clump. The situation is less clear for clump
stars, since the low-visibility branch joins the group of normal
visibility stars when $\Dnu \le 4.5\,\mu$Hz. We however notice an
overabundance of red-clump stars with low visibilities, much more
abundant than stars with visibilities above the normal value. An
example of such star is given in Fig.~\ref{fig-type2} (bottom
panel).

\paragraph{Radial mode widths}

Radial mode widths $\Gamma_0$, defined as full widths as half
maximum, were measured following the method used by
\cite{2015A&A...579A..84V}. Results are shown in
Fig.~\ref{fig-deprime}b. They are fully consistent with previous
work obtained with CoRoT and \Kepler\
\citep{2011A&A...529A..84B,2012ApJ...757..190C,2015A&A...579A..83C}
and show a clear dependence with the evolutionary stage and the
stellar mass, as will be discussed in a forthcoming paper. In the
clumps, $\Gamma_0$ of stars with depressed mixed modes behave as
for the other stars. On the RGB, these stars appear to have
slightly larger $\Gamma_0$ than the mean trend. This is however a
mass effect only: $\Gamma_0$ increases with increasing masses, and
low visibility stars show higher mass \citep{2016Natur.529..364S}.

\paragraph{Asymptotic period spacing}

Asymptotic period spacings follow the typical distribution
identified in previous work
\citep{2012A&A...540A.143M,2014A&A...572L...5M,2016A&A...588A..87V}.
We could not identify any departure to the distribution of the
$\Dnu$ -- $\Tg$ relation (Fig.~\ref{fig-deprime}c). On the RGB,
stars with depressed mixed modes show slightly lower values of
$\Tg$ than the mean case, in agreement with their mass
distribution \citep{2016A&A...588A..87V}. Values are normal in the
red clump. The large mass range and the non-degenerate conditions
for helium ignition of secondary red clump stars explains the
spread in the distribution of their seismic parameters, so that
the spread for stars with depressed modes does not allow to draw
any conclusion.

\paragraph{Coupling factors}

Coupling factors $q$ were measured by \cite{mosser} for about
4\,000 stars among the data set analyzed by
\cite{2016A&A...588A..87V}. These factors are derived from the
optimization of the method introduced by
\cite{2015A&A...584A..50M} for analyzing mixed modes. Results are
shown in Fig.~\ref{fig-deprime}d, where stars with a low
dipole-mode visibility are identified. We refer to \cite{mosser}
for the discussion of the general trends observed in $q$ as a
function of the evolutionary stage. Here, we note that stars with
depressed modes behave as the other stars. This suggests that the
extent of the evanescent region between the pressure and gravity
components is not impacted by the mechanism responsible for the
amplitude mitigation, so that it is very similar as for normal
stars \citep{1989nos..book.....U,takata}.

\paragraph{Rotation}

Fitting rotation requires a high signal-to-noise ratio in the
oscillation spectrum, so that the difficulty of this measurement
explains the limited number of stars with a complete fit. It is
evidently a bias due to the low visibilities. For the same reason,
more fits than expected are obtained for the mixed-mode patterns
of stars nearly seen pole-on, which are simpler than the general
case since the rotational multiplets are reduced to the zonal
modes (with an azimuthal order $m$=0). In such cases, the core
rotation remains undetermined. When measured, rotational
splittings show the typical distributions defined for red giants
\citep{2012A&A...548A..10M,2014A&A...564A..27D,2015A&A...580A..96D}.
\\

\begin{table}
  \begin{tabular}{lccc}
    \hline
    Evolutionary  & $\Dnu$  & $\Gamma_0$ & $\tau_0$ \\
    status        &($\mu$Hz)&    (nHz) & (day) \\
    \hline
    RGB           &  15     & 118$\pm$25 & 15.5$\pm$3.3  \\
    RGB           &  10     & 121$\pm$23 & 15.1$\pm$3.0 \\
    RGB           &  6      & 118$\pm$30 & 15.6$\pm$3.9  \\
    red clump     &  4      & 145$\pm$37 & 12.6$\pm$3.2 \\
   secondary clump&  7      & 178$\pm$45 & 10.3$\pm$2.6 \\
    \hline
  \end{tabular}
  \caption{Mean mode width and lifetime (defined as the e-folding time of the mode energy),
  depending on the evolutionary, for red giants}\label{tab-gamma}
\end{table}

\begin{figure}
\includegraphics[width=8.8cm]{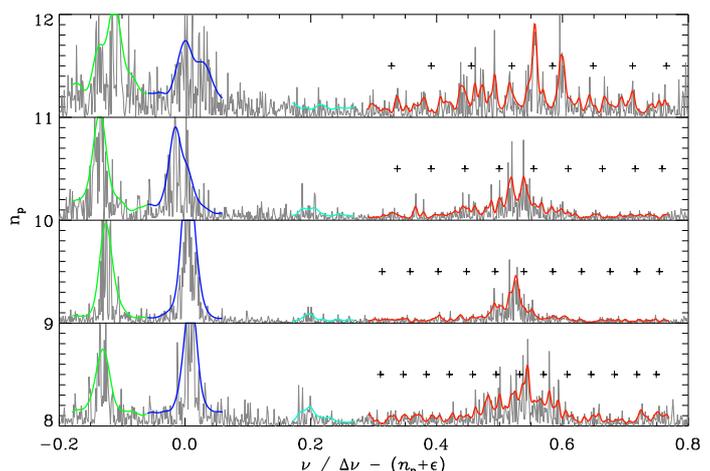}
\caption{\'Echelle spectrum of the RGB star KIC 9711269. Colored
lines, with the same color code as in Fig.~\ref{fig-type},
emphasize the structure of the modes. Plus symbols approximately
indicate the period spacings derived from the asymptotic
expansion; the fit of the mixed modes, which would imply the fit
of the rotational splittings, is however not
possible.\label{fig-structures}}
\end{figure}

\begin{figure}
\includegraphics[width=8.8cm]{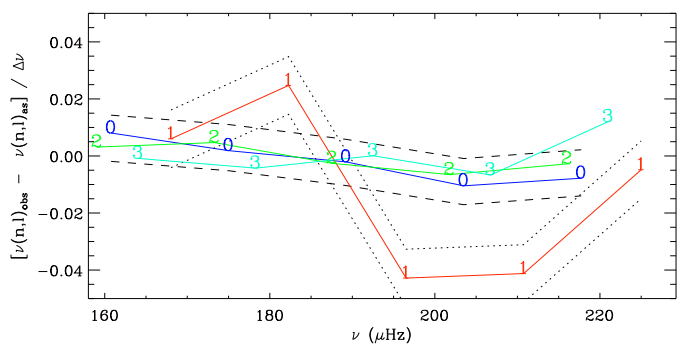}
\caption{The evidence of dipole mixed modes is provided by the
position of low-degree modes with respect to the second-order
asymptotic expansion of pure pressure modes, for the red giant
5810513 observed by \Kepler\ (Fig. \ref{fig-type}bottom). The
shifts are expressed in $\Dnu$ units. The dashed lines indicate
the region where pure pressure low-degree modes are expected; the
dotted lines provide uncertainties for the positions of the dipole
modes.\label{fig-low-dip}}
\end{figure}

\subsection{Prevalence of depressed mixed modes\label{hidden}}

As the number of stars where the mixed-mode pattern can be fitted
is limited, we checked whether the properties they display are
verified by other stars.

\subsubsection{Depressed modes \emph{versus} pure pressure modes}

A large number of oscillation spectra show peaks with a height
much above eight times the background levels. Even in the case of
low signal-to-noise ratio oscillation spectra, such peaks cannot
be all created by noise. When their identification with radial,
quadrupole or octupole modes is excluded, we must conclude that
depressed dipole mixed modes are obviously present in the whole
spectrum. An example of such a star is given in
Fig.~\ref{fig-type}b. When smoothed, oscillation spectra of such
stars exhibit the typical mixed-mode pattern
(Fig.~\ref{fig-structures}).

In other cases, mixed modes are not apparent or cannot be
distinguished from the noise (Fig.~\ref{fig-type}c). The
identification of dipole mixed modes then requires different tools
than the ones used earlier in this paper. In principle, the energy
of the dipole modes peaks at the expected position of pressure
dominated modes if modes are not mixed, but can be shifted by mode
coupling. Hence, measuring this shift provides a way to identify
whether modes are mixed or not. The measurement of the position of
the dipole modes has to fight against the acoustic glitch
\citep{2015A&A...579A..84V}, the noise induced by the background
contribution, and the intrinsic shift due to finite lifetimes. We
identified bright stars with high quality spectra on the low RGB,
where the conditions of measurement are made easier
(Fig.~\ref{fig-type}c). The shift of the actual position of the
dipole modes with respect to the asymptotic expansion is shown in
Fig.~\ref{fig-low-dip}, together with the position shifts of
radial, quadrupole and octupole modes. The curves for these modes
are remarkably close to each other, even for $\ell=3$ modes,
whereas dipole modes show a modulation as large as $\pm 0.04
\,\Dnu$.

Synthetic tests were performed to evaluate the noise contribution.
We reproduced typical conditions of observation and measured the
location of pure pressure dipole modes. In the conservative case
of a dipole mode\-width five times larger than the radial mode
width, that is much larger than the dipole width observed in other
stars, shifts were less than $0.01\ \Dnu$ away of their expected
position. The example shown in Fig.~\ref{fig-low-dip} is
representative of any spectrum with a high enough signal-to-noise
ratio. From this study, we conclude that depressed dipole mixed
modes are not an exception, but evidently the rule. Very low
visibilities cannot be associated exclusively with the full
suppression of the oscillation in the core.

\subsubsection{Asymptotic period spacings}

Asymptotic period spacings can be measured independent of the
identification of the mixed-mode pattern
\citep{2016A&A...588A..87V}. Measurements being difficult when
dipole modes are depressed, we focussed this study on all RGB
stars with a magnitude brighter than $\mV =11$, and performed an
individual analysis of each oscillation spectra. These individual
studies provided us with the measurement of the asymptotic period
spacings in more than 90\,\% of the cases. As a by-product, they
also confirmed that the structure of the excess power near the
pressure-dominated mixed modes cannot result from the simple
broadening of a single dipole pressure mode. This study fully
confirmed that asymptotic period spacings of stars with depressed
mixed modes are normal.

\subsection{Summary of the observations: depressed modes are
mixed}

From the observations, we have derived that depressed modes are
mixed and that their seismic properties, including
their asymptotic period spacings, are normal.\\
- This information was directly obtained for the  \nfit\ red
giants where the signal-to-noise ratio is high enough to fit the
mixed-mode spectrum. \\
- For \ndeprime\ stars, we measured the individual values of the
coupling factors and of the period spacings. Again, these seismic
values are normal. These stars are considered in Section
\ref{modele}.\\
- At very low visibility, estimating the asymptotic period spacing
is impossible. However, we showed that the dipole modes are
shifted with respect to the expected
position of pure pressure modes.\\
- The previous cases represent a small fraction of the \nlow\
stars with low amplitudes, but we showed that more than 90\,\% of
the stars with depressed modes brighter than $\mV=11$ on the RGB
have a mixed-mode pattern with a normal period spacing. This
prevalence can be extrapolated to fainter stars since we do not
expect any bias with magnitude. The situation for clump stars is
comparable.

\begin{figure}
\includegraphics[width=8.8cm]{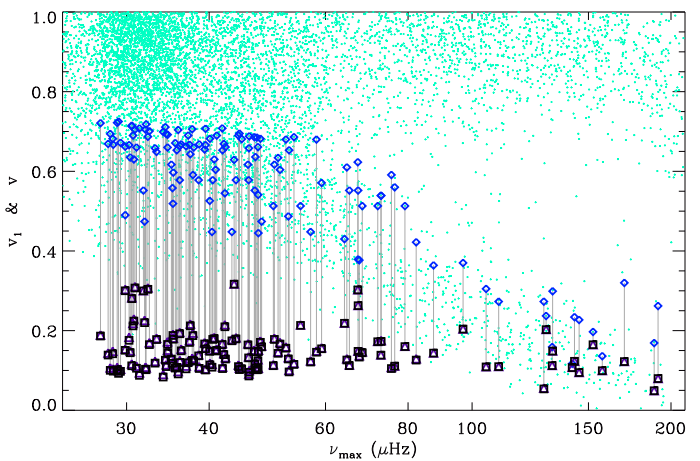}
\caption{Observed visibilities $v_1$ for the \Kepler\ public data
set of red giants, as a function of $\numax$. Values for stars
showing depressed mixed modes are emphasized with large diamonds
and compared to the computed visibilities: large dark triangles
for the low-coupling case
\protect\citep[Eq.~(\ref{eq-visibility-low-q}), as
in][]{1989nos..book.....U,2015Sci...350..423F} are very close to
black squares for the strong-coupling case
(Eq.~\ref{eq-visibility-q}). Gray lines connect the observed and
modelled values.}\label{fig-visi}
\end{figure}

\section{Observed versus predicted
visibilities\label{modele}}

As shown above, the observation of many red giants with depressed
mixed modes invalidates the hypothesis of full suppression of the
oscillation in the core for explaining low visibilities. This
opens questions on the validity of Eq.~(\ref{eq-visibility}) to
explain the low visibility. Hence, we need to check whether the
prediction of Eq.~(\ref{eq-visibility}) is sustained. As our
analysis is based on observations, we aim at checking
Eq.~(\ref{eq-visibility}) through Eq.~(\ref{eq-visibility-q}).
Therefore, we first justify the validity of using
Eq.~(\ref{eq-visibility-q}), then test various hypotheses
introduced by \cite{2015Sci...350..423F} to explain low
visibilities.

\subsection{Global seismic parameters}

Using Eq.~(\ref{eq-visibility-q}) for testing the observed
visibilities requires information on the coupling factors $q$  and
on the radial mode widths $\Gamma_0$. The set of red giants with a
low visibility and for which  all parameters of
Eq.~(\ref{eq-visibility-q}) are measured is composed of \ndeprime\
stars.  We assumed that the mechanism responsible for the extra
damping does not modify the stellar interior structure, so that
$q$ is representative of the transmission $T$. This assumption is
theoretically justified by the analysis presented in
\cite{takatab} and observationally verified: as seen above, all
stars have similar $q$ and similar $\Gamma_0$, regardless the
visibility (Figs.~\ref{fig-deprime}b and \ref{fig-deprime}d). In
that respect, the stars with depressed mixed modes provide us
indirectly with relevant tests, and Eq.~(\ref{eq-visibility-q})
can be used to test Eq.~(\ref{eq-visibility}).

\subsection{A significant disagreement}

Assuming that $q$ measured from mixed modes can replace the $T$
value (Eq.~\ref{eqt-relation-T-q}) and using observed $\Gamma_0$,
we could compare the relation between the reduced observed
visibilities $v_1$ and the calculated depressed visibility $v$
predicted by Eq.~(\ref{eq-visibility-q}). Contrary to previous
work, the estimated visibilities do not match the observed
visibilities, with modelled values significantly smaller than the
observed values (Fig.~\ref{fig-visi}). We stress that the
discrepancy is not due to the use of the formalism correct for
strong coupling introduced by \cite{takata} and \cite{takatab}. In
fact, the difference is as high as a factor of 4 for the term
$\tau_0\, T^2$ of Eq.~(\ref{eq-visibility}). Relative
uncertainties on $q$ and $\Gamma_0$ cannot explain such a high
difference. We note however that the disagreement is consistent
with partial suppression of the oscillation in the core since
observed visibilities are larger than modelled values. This
evidence is ascertained by the fact that, as made clear by
Fig.~\ref{fig-visi}, not only stars with depressed mixed modes
have observed visibilities much larger than predicted by the
model. The discrepancy certainly indicates that previous analysis
were based on inappropriate estimates of $T$.

\begin{figure}
\includegraphics[width=8.9cm]{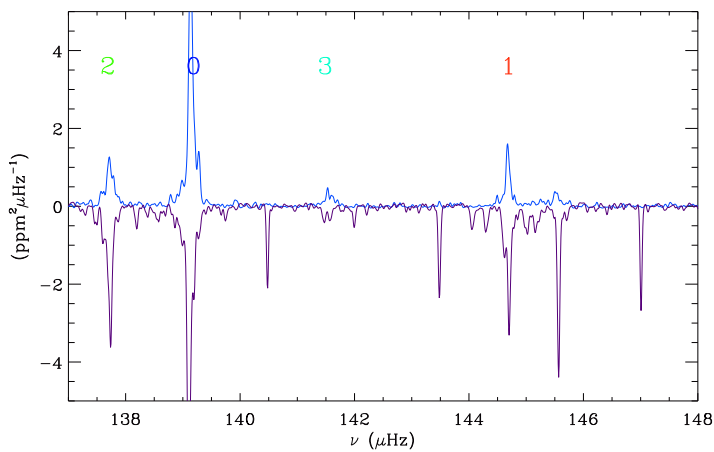}
\caption{Comparison of the smoothed spectra of two stellar twins,
KIC 5295898 (with depressed modes; blue curve) and KIC 2157650
(with normal modes, nearly seen pole on; purple curve printed
upside down). The width of the smoothing filter is 0.05\,$\mu$Hz,
much smaller than the expected width of dipole modes if not mixed.
A small shift in frequency is needed in order to superimpose the
peaks.}\label{fig-compare}
\end{figure}

\begin{figure}
\includegraphics[width=8.9cm]{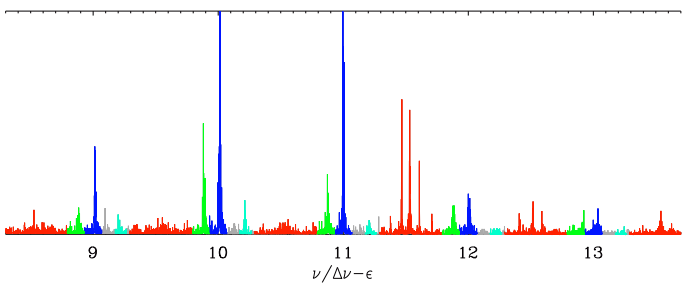}
\caption{Spectrum of the RGB star KIC 6975038 showing a strong
gradient of the dipole mode amplitudes. Dipole modes have very low
amplitudes below the radial order 11 and normal amplitudes above.
}\label{fig-gradientV1}
\end{figure}

\section{Discussion\label{discussion}}

\subsection{Depressed modes are mixed}

Important facts were inferred from the observations presented in
the previous sections: first, depressed modes are identified as
mixed modes for all stars observed with a sufficient
signal-to-noise ratio; second, period spacings of stars with
depressed modes resemble normal period spacings; third, core
rotation rates also follow the normal distribution. We next
consider these three bits of information.

\subsubsection{Depressed mixed modes}

Mixed modes were directly or indirectly observed in many stars
showing low visibilities. As already stated, the hypothesis of
full suppression of the oscillation in the core is invalidated,
since mixed modes result from the coupling of pressure waves in
the envelope and gravity waves in the core. This observational
result breaks the statement that the observation of low
visibilities implies the suppression of the oscillation in the
core. A mechanism able to only partially damp the oscillation is
needed for explaining the low visibilities.

As a result, the mechanism proposed by \cite{2015Sci...350..423F}
is not fully adequate since it relies on the total suppression of
the dipole modes in the core. Furthermore, the equivalence between
low visibility and magnetic greenhouse effect accepted in
follow-up papers \citep{2016ApJ...824...14C,2016Natur.529..364S}
is questionable.

\subsubsection{Period spacings of depressed mixed modes}

The measurement of normal period spacings in stars with depressed
modes allows us to derive further information, since such spacings
imply that the resonant cavity of gravity modes is not perturbed
by the suppression mechanism. The mechanism responsible for the
damping cannot modify the \BV\ cavity, at the high level of
precision reached with seismology.

In that respect, the scattering process associated to the magnetic
greenhouse effect is invalidated since it would modify the
resonance condition of dipole modes, with a smaller resonant
cavity for the gravity waves \citep[Fig. 1
of][]{2015Sci...350..423F}, hence larger period spacings. Larger
period spacings are not observed for depressed mixed modes. This
proves that the magnetic greenhouse effect cannot explain the many
cases where depressed mixed modes are observed. This mechanism may
work in some other cases, but proving it then requires more
information than simply the visibilities of dipole modes. In any
case, at this stage, the magnetic greenhouse effect  cannot be a
general solution for explaining depressed modes, and it is
impossible to conclude on the identity of the mechanism able to
lower the dipole mode amplitude.

\subsubsection{Rotational splittings of depressed mixed modes}

The identification of the normal rotational splittings in stars
with depressed modes provides us with similar conclusions.

If there were a strong magnetic field in the cores of the stars
with depressed modes, then \cite{2016ApJ...824...14C} predict that
an extra magnetic splitting, comparable to the gravity mode period
spacing in RGB stars, would be observed. As the core rotation
rates inferred from the mixed-mode pattern follow a similar
distribution to the reference set of stars, we can rule out the
presence of the extra, magnetic splitting.

\subsection{From individual visibilities to extra damping}

The study of the widths of dipole mixed modes can give promising
information on the way amplitude are distributed in mixed modes.
We illustrate this potential with the comparison of two twin stars
with very close $\Dnu$ and $\Tg$. The star with normal visibility
is used as a reference for the other with depressed modes. The
resemblance of their mixed-mode pattern allowed us to compare
their individual visibilities $\vm$ (Fig.~\ref{fig-compare}). The
simplifications introduced either for normal stars (Section
\ref{cas-normal}) or in the extreme case of full suppression of
the oscillation in the core (Section \ref{full_suppression}) do
not hold any more for the star with depressed modes. We have to
rewrite Eq.~(\ref{eqt-ratio-benomar}) in case of an extra damping.
For radial modes, we have
\begin{equation}\label{eqt-travail-0}
   \Gamma_0 I_0 = - {1\over \omega_0} \oint \delta W_0
   ,
\end{equation}
where $\omega_0$ is the radial frequency and the cyclic integral
represents the work during one radial oscillation. For a mixed
mode, we have an extra damping, so that
\begin{equation}\label{eqt-travail-m}
   \Gamma_\nm I_\nm
   =
   - {1\over \omega_\nm} \oint \delta W_\nm
   \simeq
   - {1\over \omega_\nm} \oint (\delta W_0 + \delta W\ind{extra} )
   ,
\end{equation}
assuming that the normal dipole work is similar to the radial
work, except for the extra damping
\citep[e.g.,][]{2009A&A...506...57D,2014ApJ...781L..29B,2014A&A...572A..11G}.
Since radial and non-radial frequencies are close to each other,
we have
\begin{equation}\label{eqt-travaux}
   \Gamma_\nm I_\nm = \Gamma_0 I_0\, (1+x),
\end{equation}
with a similar definition for $x$ as in Eq.~(\ref{cas_fuller2}):
$x$ represents the relative contribution of the extra damping.

From Eq.~(\ref{eqt-ratio-benomar}), we obtain a new expression for
the mixed mode visibility,
\begin{equation}\label{eqt-mesure-dampingv}
   \vm
   \simeq {I_0 \over (1+x)\, I_\nm\ }
   \simeq {\Gamma_\nm \over (1+x)^2\ \Gamma_0}
   ,
\end{equation}
which demonstrates the capability of mixed modes to measure the
relative extra damping $x$.

If we simply assume that $x$ has limited variation in frequency,
the total contribution $v$ of the individual visibilities provides
an estimate of the extra damping,
\begin{equation}\label{eqt-mesure-dampingv-p}
   x \simeq {1 - v \over v}
   .
\end{equation}
According to this relation, the magnitude of this relative
extra-damping significantly decreases with stellar evolution. Very
low visibilities observed for stars on the low RGB are due to a
large absorption ($x\simeq 9$ when $v_1=0.1$), but $x \simeq 0.7$
only when $v_1=0.6$. For KIC 5295898 shown in
Fig.~\ref{fig-compare}, $x\simeq 2.6$.

\subsection{No man's land}

For completeness, we fitted the stars in the no man's land between
normal and low-visibility stars (Figs.~\ref{fig-deprime}a and
\ref{fig-visi}). We checked that these stars behave as other
stars, with similar seismic parameters. The presence of such stars
is crucial for at least two reasons. First, they show that
intermediate values between normal and low visibilities are
possible. Second, those stars represent the intermediate case
between normal and low visibility. This reinforces the fact that
stars where depressed mixed modes could be fully characterized are
representative of all stars with low-amplitude dipole modes.

\subsection{Visibility gradient\label{section-visibility}}

Three stars of our data set exhibit a clear visibility gradient:
KIC 6975038 (Fig.~\ref{fig-gradientV1}), 7746983, and 8561221.

- The case of KIC 6975038 is investigated in \cite{mosser}. This
star shows atypical seismic parameters: $\Tg$ is very low and $q$
is unusually large compared to the general trend on the RGB
\citep{2015A&A...584A..50M,2016A&A...588A..87V}. This star
deserves a precise modelling beyond the scope of this work.

- KIC 8561221 was identified by \cite{2014A&A...563A..84G} as the
least evolved observed star with depressed dipole modes among red
giants observed with \Kepler. It shows a very low dipole-mode
visibility (Table~\ref{tab-properties}). However, mixed modes can
be firmly identified. The asymptotic period spacing $\Tg\simeq
114.8\,$s is typical for $\Dnu\simeq 29.8\,\mu$Hz, but the core
rotation of this star seems very high. We measure a core rotation
rate of about 2.6\,$\mu$Hz, in contradiction with the values
extracted from $\ell=2$ and $\ell=3$ modes by
\cite{2014A&A...563A..84G}.

- Compared to the two previous stars, the seismic parameters of
KIC 7746983 are close to the values obtained on the RGB; only the
dipole mode visibility is atypical. This gradient appears to be
helpful for characterizing the mixed mode pattern: after KIC
8561221, this star is the second least evolved in our data set
with low-dipole visibility.\\

The change of visibility with frequency was used by
\cite{2015Sci...350..423F} as a further argument in favor of a
large magnetic field for explaining the suppression of the
oscillation, since the variation of the visibility with frequency
matches their prediction. A more conservative analysis consists in
remarking that the physics of oscillation damping has to be
frequency dependent. Red giants showing a gradient of dipole
visibility are certainly useful benchmark stars for understanding
the nature and the physics of the extra damping of the
oscillation.

\section{Conclusion\label{conclusion}}

We performed a thorough study of red giants showing low
dipole-mode visibility, based on the identification of their
dipole mode pattern and on the characterization of their global
seismic properties. We have shown that these stars share the same
global seismic parameters as other stars, regardless the value of
the dipole mode visibilities. This analysis sustains the fact that
the mechanism responsible for the damping does not significantly
impact the stellar structure and does not change the property of
the cavity where gravity waves propagate.

We were able to determine that dipole depressed modes are mixed,
even at very low visibilities. The existence of these depressed
mixed modes implies that oscillations cannot be fully suppressed
in the radiative core. We also note that the observed visibilities
are significantly higher than predicted from the modelling
assuming full suppression in the core, which is consistent with
partial suppression of the core oscillation only. Furthermore, the
observations of normal period spacings in stars with depressed
mixed modes indicates that the radiative core of these stars is
not affected by the suppression mechanism.

These precise seismic signatures indicate that the magnetic
greenhouse effect cannot explain the observed low visibilities of
dipole modes \citep{2015Sci...350..423F}. This effect supposes the
full suppression of the oscillation, which is discarded by the
fact that depressed modes are mixed. Even if the mechanism could
work with partial suppression only, the scattering process induced
by the magnetic field in the radiative core is dismissed by the
observation of the period spacings. As a result, inferring high
magnetic fields in red giant from low visibilities
\citep{2016Natur.529..364S,2016ApJ...824...14C} is at least
premature. This conclusion applies in the vast majority of stars
that show low visibilities.

The low integrated visibilities reflect an extra mode damping but,
at this stage, carry no direct information on the nature of this
damping. Another damping mechanism must be found. This mechanism,
which partially damps the dipole mixed modes, could be
characterized by the measurement of the mixed mode widths.

\begin{acknowledgements}
We acknowledge the entire \emph{Kepler} team, whose efforts made
these results possible. BM thanks Jim Fuller for interesting
discussions. BM, KB, CP, CB, MJG and RS acknowledge financial
support from the Programme National de Physique Stellaire
(CNRS/INSU), from the French space agency CNES, and from the ANR
program IDEE Interaction Des \'Etoiles et des Exoplan\`etes. MT is
partially supported by JSPS KAKENHI Grant Number 26400219. MV
acknowledges funding by the Portuguese Science foundation through
the grant CIAAUP-03/2016-BPD, in the context of the project
FIS/04434, co-funded by FEDER through the program COMPETE.

\end{acknowledgements}

\bibliographystyle{aa} 

\begin{thebibliography}{51}

\bibitem[{{Ballot} {et~al.}(2011){Ballot}, {Barban}, \& {van't
  Veer-Menneret}}]{2011A&A...531A.124B}
{Ballot}, J., {Barban}, C., \& {van't Veer-Menneret}, C. 2011,
\aap, 531, A124

\bibitem[{{Baudin} {et~al.}(2011){Baudin}, {Barban}, {Belkacem}, {Hekker},
  {Morel}, {Samadi}, {Benomar}, {Goupil}, {Carrier}, {Ballot}, {Deheuvels}, {De
  Ridder}, {Hatzes}, {Kallinger}, \& {Weiss}}]{2011A&A...529A..84B}
{Baudin}, F., {Barban}, C., {Belkacem}, K., {et~al.} 2011, \aap,
529, A84

\bibitem[{{Beck} {et~al.}(2012){Beck}, {Montalban}, {Kallinger}, {De Ridder},
  {Aerts}, {Garc{\'{\i}}a}, {Hekker}, {Dupret}, {Mosser}, {Eggenberger},
  {Stello}, {Elsworth}, {Frandsen}, {Carrier}, {Hillen}, {Gruberbauer},
  {Christensen-Dalsgaard}, {Miglio}, {Valentini}, {Bedding}, {Kjeldsen},
  {Girouard}, {Hall}, \& {Ibrahim}}]{2012Natur.481...55B}
{Beck}, P.~G., {Montalban}, J., {Kallinger}, T., {et~al.} 2012,
\nat, 481, 55

\bibitem[{{Bedding} {et~al.}(2011){Bedding}, {Mosser}, {Huber},
  {Montalb{\'a}n}, {Beck}, {Christensen-Dalsgaard}, {Elsworth},
  {Garc{\'{\i}}a}, {Miglio}, {Stello}, {White}, {De Ridder}, {Hekker}, {Aerts},
  {Barban}, {Belkacem}, {Broomhall}, {Brown}, {Buzasi}, {Carrier}, {Chaplin},
  {di Mauro}, {Dupret}, {Frandsen}, {Gilliland}, {Goupil}, {Jenkins},
  {Kallinger}, {Kawaler}, {Kjeldsen}, {Mathur}, {Noels}, {Aguirre}, \&
  {Ventura}}]{2011Natur.471..608B}
{Bedding}, T.~R., {Mosser}, B., {Huber}, D., {et~al.} 2011, \nat,
471, 608

\bibitem[{{Belkacem} {et~al.}(2015){Belkacem}, {Marques}, {Goupil}, {Mosser},
  {Sonoi}, {Ouazzani}, {Dupret}, {Mathis}, \& {Grosjean}}]{2015A&A...579A..31B}
{Belkacem}, K., {Marques}, J.~P., {Goupil}, M.~J., {et~al.} 2015,
\aap, 579,
  A31

\bibitem[{{Belkacem} {et~al.}(2008){Belkacem}, {Samadi}, {Goupil}, \&
  {Dupret}}]{2008A&A...478..163B}
{Belkacem}, K., {Samadi}, R., {Goupil}, M.-J., \& {Dupret}, M.-A.
2008, \aap,
  478, 163

\bibitem[{{Benomar} {et~al.}(2014){Benomar}, {Belkacem}, {Bedding}, {Stello},
  {Di Mauro}, {Ventura}, {Mosser}, {Goupil}, {Samadi}, \&
  {Garcia}}]{2014ApJ...781L..29B}
{Benomar}, O., {Belkacem}, K., {Bedding}, T.~R., {et~al.} 2014,
\apjl, 781, L29

\bibitem[{{Cantiello} {et~al.}(2016){Cantiello}, {Fuller}, \&
  {Bildsten}}]{2016ApJ...824...14C}
{Cantiello}, M., {Fuller}, J., \& {Bildsten}, L. 2016, \apj, 824,
14

\bibitem[{{Chaplin} {et~al.}(2011){Chaplin}, {Kjeldsen},
  {Christensen-Dalsgaard}, {Basu}, {Miglio}, {Appourchaux}, {Bedding},
  {Elsworth}, {Garc{\'{\i}}a}, {Gilliland}, {Girardi}, {Houdek}, {Karoff},
  {Kawaler}, {Metcalfe}, {Molenda-{\.Z}akowicz}, {Monteiro}, {Thompson},
  {Verner}, {Ballot}, {Bonanno}, {Brand{\~a}o}, {Broomhall}, {Bruntt},
  {Campante}, {Corsaro}, {Creevey}, {Do{\u g}an}, {Esch}, {Gai}, {Gaulme},
  {Hale}, {Handberg}, {Hekker}, {Huber}, {Jim{\'e}nez}, {Mathur}, {Mazumdar},
  {Mosser}, {New}, {Pinsonneault}, {Pricopi}, {Quirion}, {R{\'e}gulo},
  {Salabert}, {Serenelli}, {Aguirre}, {Sousa}, {Stello}, {Stevens}, {Suran},
  {Uytterhoeven}, {White}, {Borucki}, {Brown}, {Jenkins}, {Kinemuchi}, {Van
  Cleve}, \& {Klaus}}]{2011Sci...332..213C}
{Chaplin}, W.~J., {Kjeldsen}, H., {Christensen-Dalsgaard}, J.,
{et~al.} 2011,
  Science, 332, 213

\bibitem[{{Corsaro} {et~al.}(2015){Corsaro}, {De Ridder}, \&
  {Garc{\'{\i}}a}}]{2015A&A...579A..83C}
{Corsaro}, E., {De Ridder}, J., \& {Garc{\'{\i}}a}, R.~A. 2015,
\aap, 579, A83

\bibitem[{{Corsaro} {et~al.}(2012){Corsaro}, {Stello}, {Huber}, {Bedding},
  {Bonanno}, {Brogaard}, {Kallinger}, {Benomar}, {White}, {Mosser}, {Basu},
  {Chaplin}, {Christensen-Dalsgaard}, {Elsworth}, {Garc{\'{\i}}a}, {Hekker},
  {Kjeldsen}, {Mathur}, {Meibom}, {Hall}, {Ibrahim}, \&
  {Klaus}}]{2012ApJ...757..190C}
{Corsaro}, E., {Stello}, D., {Huber}, D., {et~al.} 2012, \apj,
757, 190

\bibitem[{{De Ridder} {et~al.}(2009){De Ridder}, {Barban}, {Baudin}, {Carrier},
  {Hatzes}, {Hekker}, {Kallinger}, {Weiss}, {Baglin}, {Auvergne}, {Samadi},
  {Barge}, \& {Deleuil}}]{2009Natur.459..398D}
{De Ridder}, J., {Barban}, C., {Baudin}, F., {et~al.} 2009, \nat,
459, 398

\bibitem[{{Deheuvels} {et~al.}(2015){Deheuvels}, {Ballot}, {Beck}, {Mosser},
  {{\O}stensen}, {Garc{\'{\i}}a}, \& {Goupil}}]{2015A&A...580A..96D}
{Deheuvels}, S., {Ballot}, J., {Beck}, P.~G., {et~al.} 2015, \aap,
580, A96

\bibitem[{{Deheuvels} {et~al.}(2014){Deheuvels}, {Do{\u g}an}, {Goupil},
  {Appourchaux}, {Benomar}, {Bruntt}, {Campante}, {Casagrande}, {Ceillier},
  {Davies}, {De Cat}, {Fu}, {Garc{\'{\i}}a}, {Lobel}, {Mosser}, {Reese},
  {Regulo}, {Schou}, {Stahn}, {Thygesen}, {Yang}, {Chaplin},
  {Christensen-Dalsgaard}, {Eggenberger}, {Gizon}, {Mathis},
  {Molenda-{\.Z}akowicz}, \& {Pinsonneault}}]{2014A&A...564A..27D}
{Deheuvels}, S., {Do{\u g}an}, G., {Goupil}, M.~J., {et~al.} 2014,
\aap, 564,
  A27

\bibitem[{{Deheuvels} {et~al.}(2012){Deheuvels}, {Garc{\'{\i}}a}, {Chaplin},
  {Basu}, {Antia}, {Appourchaux}, {Benomar}, {Davies}, {Elsworth}, {Gizon},
  {Goupil}, {Reese}, {Regulo}, {Schou}, {Stahn}, {Casagrande},
  {Christensen-Dalsgaard}, {Fischer}, {Hekker}, {Kjeldsen}, {Mathur}, {Mosser},
  {Pinsonneault}, {Valenti}, {Christiansen}, {Kinemuchi}, \&
  {Mullally}}]{2012ApJ...756...19D}
{Deheuvels}, S., {Garc{\'{\i}}a}, R.~A., {Chaplin}, W.~J.,
{et~al.} 2012, \apj,
  756, 19

\bibitem[{{Dupret} {et~al.}(2009){Dupret}, {Belkacem}, {Samadi}, {Montalban},
  {Moreira}, {Miglio}, {Godart}, {Ventura}, {Ludwig}, {Grigahc{\`e}ne},
  {Goupil}, {Noels}, \& {Caffau}}]{2009A&A...506...57D}
{Dupret}, M., {Belkacem}, K., {Samadi}, R., {et~al.} 2009, \aap,
506, 57

\bibitem[{{Dziembowski}(2012)}]{2012A&A...539A..83D}
{Dziembowski}, W.~A. 2012, \aap, 539, A83

\bibitem[{{Epstein} {et~al.}(2014){Epstein}, {Elsworth}, {Johnson}, {Shetrone},
  {Mosser}, {Hekker}, {Tayar}, {Harding}, {Pinsonneault}, {Silva Aguirre},
  {Basu}, {Beers}, {Bizyaev}, {Bedding}, {Chaplin}, {Frinchaboy},
  {Garc{\'{\i}}a}, {Garc{\'{\i}}a P{\'e}rez}, {Hearty}, {Huber}, {Ivans},
  {Majewski}, {Mathur}, {Nidever}, {Serenelli}, {Schiavon}, {Schneider},
  {Sch{\"o}nrich}, {Sobeck}, {Stassun}, {Stello}, \&
  {Zasowski}}]{2014ApJ...785L..28E}
{Epstein}, C.~R., {Elsworth}, Y.~P., {Johnson}, J.~A., {et~al.}
2014, \apjl,
  785, L28

\bibitem[{{Fuller} {et~al.}(2015){Fuller}, {Cantiello}, {Stello}, {Garcia}, \&
  {Bildsten}}]{2015Sci...350..423F}
{Fuller}, J., {Cantiello}, M., {Stello}, D., {Garcia}, R.~A., \&
{Bildsten}, L.
  2015, Science, 350, 423

\bibitem[{{Garc{\'{\i}}a} {et~al.}(2014){Garc{\'{\i}}a}, {P{\'e}rez
  Hern{\'a}ndez}, {Benomar}, {Silva Aguirre}, {Ballot}, {Davies}, {Do{\u g}an},
  {Stello}, {Christensen-Dalsgaard}, {Houdek}, {Ligni{\`e}res}, {Mathur},
  {Takata}, {Ceillier}, {Chaplin}, {Mathis}, {Mosser}, {Ouazzani},
  {Pinsonneault}, {Reese}, {R{\'e}gulo}, {Salabert}, {Thompson}, {van Saders},
  {Neiner}, \& {De Ridder}}]{2014A&A...563A..84G}
{Garc{\'{\i}}a}, R.~A., {P{\'e}rez Hern{\'a}ndez}, F., {Benomar},
O., {et~al.}
  2014, \aap, 563, A84

\bibitem[{{Gaulme} {et~al.}(2014){Gaulme}, {Jackiewicz}, {Appourchaux}, \&
  {Mosser}}]{2014ApJ...785....5G}
{Gaulme}, P., {Jackiewicz}, J., {Appourchaux}, T., \& {Mosser}, B.
2014, \apj,
  785, 5

\bibitem[{{Goupil} {et~al.}(2013){Goupil}, {Mosser}, {Marques}, {Ouazzani},
  {Belkacem}, {Lebreton}, \& {Samadi}}]{2013A&A...549A..75G}
{Goupil}, M.~J., {Mosser}, B., {Marques}, J.~P., {et~al.} 2013,
\aap, 549, A75

\bibitem[{{Grosjean} {et~al.}(2014){Grosjean}, {Dupret}, {Belkacem},
  {Montalban}, {Samadi}, \& {Mosser}}]{2014A&A...572A..11G}
{Grosjean}, M., {Dupret}, M.-A., {Belkacem}, K., {et~al.} 2014,
\aap, 572, A11

\bibitem[{{Huber} {et~al.}(2014){Huber}, {Silva Aguirre}, {Matthews},
  {Pinsonneault}, {Gaidos}, {Garc{\'{\i}}a}, {Hekker}, {Mathur}, {Mosser},
  {Torres}, {Bastien}, {Basu}, {Bedding}, {Chaplin}, {Demory}, {Fleming},
  {Guo}, {Mann}, {Rowe}, {Serenelli}, {Smith}, \&
  {Stello}}]{2014ApJS..211....2H}
{Huber}, D., {Silva Aguirre}, V., {Matthews}, J.~M., {et~al.}
2014, \apjs, 211,
  2

\bibitem[{{Kallinger} {et~al.}(2012){Kallinger}, {Hekker}, {Mosser}, {De
  Ridder}, {Bedding}, {Elsworth}, {Gruberbauer}, {Guenther}, {Stello}, {Basu},
  {Garc{\'{\i}}a}, {Chaplin}, {Mullally}, {Still}, \&
  {Thompson}}]{2012A&A...541A..51K}
{Kallinger}, T., {Hekker}, S., {Mosser}, B., {et~al.} 2012, \aap,
541, A51

\bibitem[{{Kallinger} {et~al.}(2010){Kallinger}, {Mosser}, {Hekker}, {Huber},
  {Stello}, {Mathur}, {Basu}, {Bedding}, {Chaplin}, {De Ridder}, {Elsworth},
  {Frandsen}, {Garc{\'{\i}}a}, {Gruberbauer}, {Matthews}, {Borucki}, {Bruntt},
  {Christensen-Dalsgaard}, {Gilliland}, {Kjeldsen}, \&
  {Koch}}]{2010A&A...522A...1K}
{Kallinger}, T., {Mosser}, B., {Hekker}, S., {et~al.} 2010, \aap,
522, A1

\bibitem[{{Lagarde} {et~al.}(2016){Lagarde}, {Bossini}, {Miglio}, {Vrard}, \&
  {Mosser}}]{2016MNRAS.457L..59L}
{Lagarde}, N., {Bossini}, D., {Miglio}, A., {Vrard}, M., \&
{Mosser}, B. 2016,
  \mnras, 457, L59

\bibitem[{{Mathur} {et~al.}(2011){Mathur}, {Hekker}, {Trampedach}, {Ballot},
  {Kallinger}, {Buzasi}, {Garc{\'{\i}}a}, {Huber}, {Jim{\'e}nez}, {Mosser},
  {Bedding}, {Elsworth}, {R{\'e}gulo}, {Stello}, {Chaplin}, {De Ridder},
  {Hale}, {Kinemuchi}, {Kjeldsen}, {Mullally}, \&
  {Thompson}}]{2011ApJ...741..119M}
{Mathur}, S., {Hekker}, S., {Trampedach}, R., {et~al.} 2011, \apj,
741, 119

\bibitem[{{Michel} {et~al.}(2008){Michel}, {Baglin}, {Auvergne}, {Catala},
  {Samadi}, {Baudin}, {Appourchaux}, {Barban}, {Weiss}, {Berthomieu},
  {Boumier}, {Dupret}, {Garcia}, {Fridlund}, {Garrido}, {Goupil}, {Kjeldsen},
  {Lebreton}, {Mosser}, {Grotsch-Noels}, {Janot-Pacheco}, {Provost},
  {Roxburgh}, {Thoul}, {Toutain}, {Tiph{\`e}ne}, {Turck-Chieze}, {Vauclair},
  {Vauclair}, {Aerts}, {Alecian}, {Ballot}, {Charpinet}, {Hubert},
  {Ligni{\`e}res}, {Mathias}, {Monteiro}, {Neiner}, {Poretti}, {Renan de
  Medeiros}, {Ribas}, {Rieutord}, {Cort{\'e}s}, \&
  {Zwintz}}]{2008Sci...322..558M}
{Michel}, E., {Baglin}, A., {Auvergne}, M., {et~al.} 2008,
Science, 322, 558

\bibitem[{{Miglio} {et~al.}(2009){Miglio}, {Montalb{\'a}n}, {Baudin},
  {Eggenberger}, {Noels}, {Hekker}, {De Ridder}, {Weiss}, \&
  {Baglin}}]{2009A&A...503L..21M}
{Miglio}, A., {Montalb{\'a}n}, J., {Baudin}, F., {et~al.} 2009,
\aap, 503, L21

\bibitem[{{Montalb{\'a}n} \& {Noels}(2013)}]{2013EPJWC..4303002M}
{Montalb{\'a}n}, J. \& {Noels}, A. 2013, in European Physical
Journal Web of
  Conferences, Vol.~43, European Physical Journal Web of Conferences, 3002

\bibitem[{{Mosser}(2015)}]{2015EAS....73....3M}
{Mosser}, B. 2015, in EAS Publications Series, Vol.~73, EAS
Publications
  Series, 3--110

\bibitem[{{Mosser} {et~al.}(2011){Mosser}, {Barban}, {Montalb{\'a}n}, {Beck},
  {Miglio}, {Belkacem}, {Goupil}, {Hekker}, {De Ridder}, {Dupret}, {Elsworth},
  {Noels}, {Baudin}, {Michel}, {Samadi}, {Auvergne}, {Baglin}, \&
  {Catala}}]{2011A&A...532A..86M}
{Mosser}, B., {Barban}, C., {Montalb{\'a}n}, J., {et~al.} 2011,
\aap, 532, A86

\bibitem[{{Mosser} {et~al.}(2010){Mosser}, {Belkacem}, {Goupil}, {Miglio},
  {Morel}, {Barban}, {Baudin}, {Hekker}, {Samadi}, {De Ridder}, {Weiss},
  {Auvergne}, \& {Baglin}}]{2010A&A...517A..22M}
{Mosser}, B., {Belkacem}, K., {Goupil}, M., {et~al.} 2010, \aap,
517, A22

\bibitem[{{Mosser} {et~al.}(2014){Mosser}, {Benomar}, {Belkacem}, {Goupil},
  {Lagarde}, {Michel}, {Lebreton}, {Stello}, {Vrard}, {Barban}, {Bedding},
  {Deheuvels}, {Chaplin}, {De Ridder}, {Elsworth}, {Montalban}, {Noels},
  {Ouazzani}, {Samadi}, {White}, \& {Kjeldsen}}]{2014A&A...572L...5M}
{Mosser}, B., {Benomar}, O., {Belkacem}, K., {et~al.} 2014, \aap,
572, L5

\bibitem[{{Mosser} {et~al.}(2012{\natexlab{a}}){Mosser}, {Elsworth}, {Hekker},
  {Huber}, {Kallinger}, {Mathur}, {Belkacem}, {Goupil}, {Samadi}, {Barban},
  {Bedding}, {Chaplin}, {Garc{\'{\i}}a}, {Stello}, {De Ridder}, {Middour},
  {Morris}, \& {Quintana}}]{2012A&A...537A..30M}
{Mosser}, B., {Elsworth}, Y., {Hekker}, S., {et~al.}
2012{\natexlab{a}}, \aap,
  537, A30

\bibitem[{{Mosser} {et~al.}(2012{\natexlab{b}}){Mosser}, {Goupil}, {Belkacem},
  {Marques}, {Beck}, {Bloemen}, {De Ridder}, {Barban}, {Deheuvels}, {Elsworth},
  {Hekker}, {Kallinger}, {Ouazzani}, {Pinsonneault}, {Samadi}, {Stello},
  {Garc{\'{\i}}a}, {Klaus}, {Li}, {Mathur}, \& {Morris}}]{2012A&A...548A..10M}
{Mosser}, B., {Goupil}, M.~J., {Belkacem}, K., {et~al.}
2012{\natexlab{b}},
  \aap, 548, A10

\bibitem[{{Mosser} {et~al.}(2012{\natexlab{c}}){Mosser}, {Goupil}, {Belkacem},
  {Michel}, {Stello}, {Marques}, {Elsworth}, {Barban}, {Beck}, {Bedding}, {De
  Ridder}, {Garc{\'{\i}}a}, {Hekker}, {Kallinger}, {Samadi}, {Stumpe},
  {Barclay}, \& {Burke}}]{2012A&A...540A.143M}
{Mosser}, B., {Goupil}, M.~J., {Belkacem}, K., {et~al.}
2012{\natexlab{c}},
  \aap, 540, A143

\bibitem[{{Mosser} {et~al.}(2013){Mosser}, {Michel}, {Belkacem}, {Goupil},
  {Baglin}, {Barban}, {Provost}, {Samadi}, {Auvergne}, \&
  {Catala}}]{2013A&A...550A.126M}
{Mosser}, B., {Michel}, E., {Belkacem}, K., {et~al.} 2013, \aap,
550, A126

\bibitem[{{Mosser} \& {Miglio}(2016)}]{2016cole.book..197M}
{Mosser}, B. \& {Miglio}, A. 2016, {IV.2 Pulsating red giant
stars}, ed. {CoRot
  Team}, 197

\bibitem[{{Mosser} {et~al.}(2016){Mosser}, {Pin\c con}, {Vrard}, {Belkacem},
  {Masao}, {Masao}, \& {Masao}}]{mosser}
{Mosser}, B., {Pin\c con}, C., {Vrard}, M., {et~al.} 2016, in
preparation

\bibitem[{{Mosser} {et~al.}(2015){Mosser}, {Vrard}, {Belkacem}, {Deheuvels}, \&
  {Goupil}}]{2015A&A...584A..50M}
{Mosser}, B., {Vrard}, M., {Belkacem}, K., {Deheuvels}, S., \&
{Goupil}, M.~J.
  2015, \aap, 584, A50

\bibitem[{{Samadi} {et~al.}(2015){Samadi}, {Belkacem}, \&
  {Sonoi}}]{2015EAS....73..111S}
{Samadi}, R., {Belkacem}, K., \& {Sonoi}, T. 2015, in EAS
Publications Series,
  Vol.~73, EAS Publications Series, 111--191

\bibitem[{{Stello} {et~al.}(2016{\natexlab{a}}){Stello}, {Cantiello}, {Fuller},
  {Garcia}, \& {Huber}}]{2016PASA...33...11S}
{Stello}, D., {Cantiello}, M., {Fuller}, J., {Garcia}, R.~A., \&
{Huber}, D.
  2016{\natexlab{a}}, \pasa, 33, e011

\bibitem[{{Stello} {et~al.}(2016{\natexlab{b}}){Stello}, {Cantiello}, {Fuller},
  {Huber}, {Garc{\'{\i}}a}, {Bedding}, {Bildsten}, \&
  {Aguirre}}]{2016Natur.529..364S}
{Stello}, D., {Cantiello}, M., {Fuller}, J., {et~al.}
2016{\natexlab{b}}, \nat,
  529, 364

\bibitem[{{Stello} {et~al.}(2009){Stello}, {Chaplin}, {Basu}, {Elsworth}, \&
  {Bedding}}]{2009MNRAS.400L..80S}
{Stello}, D., {Chaplin}, W.~J., {Basu}, S., {Elsworth}, Y., \&
{Bedding}, T.~R.
  2009, \mnras, 400, L80

\bibitem[{{Takata}(2016{\natexlab{a}})}]{takata}
{Takata}, M. 2016{\natexlab{a}}, PASJ, in press

\bibitem[{{Takata}(2016{\natexlab{b}})}]{takatab}
{Takata}, M. 2016{\natexlab{b}}, \pasj, in press

\bibitem[{{Unno} {et~al.}(1989){Unno}, {Osaki}, {Ando}, {Saio}, \&
  {Shibahashi}}]{1989nos..book.....U}
{Unno}, W., {Osaki}, Y., {Ando}, H., {Saio}, H., \& {Shibahashi},
H. 1989,
  {Nonradial oscillations of stars}, ed. {Unno, W., Osaki, Y., Ando, H., Saio,
  H., \& Shibahashi, H.}

\bibitem[{{Vrard} {et~al.}(2015){Vrard}, {Mosser}, {Barban}, {Belkacem},
  {Elsworth}, {Kallinger}, {Hekker}, {Samadi}, \& {Beck}}]{2015A&A...579A..84V}
{Vrard}, M., {Mosser}, B., {Barban}, C., {et~al.} 2015, \aap, 579,
A84

\bibitem[{{Vrard} {et~al.}(2016){Vrard}, {Mosser}, \&
  {Samadi}}]{2016A&A...588A..87V}
{Vrard}, M., {Mosser}, B., \& {Samadi}, R. 2016, \aap, 588, A87

\end{thebibliography}

\end{document}